\documentclass[useAMS,usenatbib]{mn2e}

\usepackage{graphicx}
\usepackage{natbib}
\usepackage{subfigure}
   \usepackage{amsmath}
   \usepackage{amsfonts}   
   \usepackage{amssymb}    
\usepackage{longtable,lscape}
\usepackage[T1]{fontenc}
\usepackage{aecompl}
\usepackage{pslatex}

\def\aap{A\&A}
\def\apjl{ApJL}
\def\apjs{ApJS}

\title[The nature of H$\alpha$ emitters at $z \sim 2.23$]{On the nature of H$\alpha$ emitters at $z \sim 2$ from the HiZELS survey: physical properties, Ly$\alpha$ escape fraction, and main sequence \thanks{{\it Herschel} is an ESA space observatory with science instruments provided by European-led Principal Investigator consortia and with important participation from NASA.}}
\author[I. Oteo et al.]
{\parbox{\textwidth}{I. Oteo$^{1,2}$ \thanks{E-mail: \texttt{ivanoteogomez@gmail.com}},
D. Sobral$^{3,4,5}$, 
R. J. Ivison$^{1,2}$, 
I. Smail$^{6}$,
P. N. Best$^{1}$,
J. Cepa$^{7,8}$, and
A. M P\'erez-Garc\'ia$^{7,8}$
}\vspace{0.4cm}\\
$^{1}$SUPA, Institute for Astronomy, University of Edinburgh, Royal Observatory, Blackford Hill, Edinburgh EH9 3HJ \\
$^{2}$European Southern Observatory, Karl-Schwarzschild-Str. 2, 85748 Garching, Germany\\
$^{3}$Instituto de Astrof\'isica e Ci\^encias do Espa\c co, Universidade de Lisboa, OAL, Tapada da Ajuda, PT1349-018 Lisboa, Portugal\\
$^{4}$Departamento de F\'{i}sica, Faculdade de Ci\^{e}ncias, Universidade de Lisboa, Edif\'{i}cio C8, Campo Grande, PT1749-016 Lisbon, Portugal \\
$^{5}$Leiden Observatory, Leiden University, PO Box 9513, NL-2300 RA Leiden, the Netherlands\\
$^{6}$Institute for Computational Cosmology, Durham University, South Road, Durham DH1 3LE, UK\\
$^{7}$Instituto de Astrof{\'i}sica de Canarias (IAC), E-38200 La Laguna, Tenerife, Spain\\
$^{8}$Departamento de Astrof{\'i}sica, Universidad de La Laguna (ULL), E-38205 La Laguna, Tenerife, Spain\\
}

\begin{document}

\date{Accepted ??. Received ??; in original form ??}

\pagerange{\pageref{firstpage}--\pageref{lastpage}} \pubyear{2002}

\maketitle

\label{firstpage}

\begin{abstract}  

We present a detailed multi-wavelength study (from rest-frame UV to far-IR) of narrow-band (NB) selected, star-forming (SF) H$\alpha$ emitters (HAEs) at $z \sim 2.23$ taken from the High Redshift(Z) Emission Line Survey (HiZELS). We find that HAEs have similar SED-derived properties and colors to $sBzK$ galaxies and probe a well-defined portion of the SF population at $z \sim 2$. This is not true for Ly$\alpha$ emitters (LAEs), which are strongly biased towards blue, less massive galaxies (missing a significant percentage of the SF population). Combining our H$\alpha$ observations with matched, existing Ly$\alpha$ data we determine that the Ly$\alpha$ escape fraction ($f_{\rm esc}$) is low (only $\sim$ 4.5\% of HAEs show Ly$\alpha$ emission) and decreases with increasing dust attenuation, UV continuum slope, stellar mass, and star formation rate (SFR). This suggests that Ly$\alpha$ preferentially escapes from blue galaxies with low dust attenuation. However, a small population of red and massive LAEs is also present in agreement with previous works. This indicates that dust and Ly$\alpha$ are not mutually exclusive. Using different and completely independent measures of the total SFR we show that the H$\alpha$ emission is an excellent tracer of star formation at $z \sim 2$ with deviations typically lower than 0.3 dex for individual galaxies. We find that the slope and zero-point of the HAE main-sequence (MS) at $z \sim 2$ strongly depend on the dust correction method used to recover SFR, although they are consistent with previous works when similar assumptions are made.
\end{abstract}



\section{Introduction}\label{intro}

A wide variety of multi-wavelength surveys indicate that the rate at which galaxies form stars has changed with cosmic time, increasing by about one order of magnitude from $z \sim 0$ to $z \sim 1-2$, when the cosmic star formation had a likely maximum \citep{Lilly1996ApJ...460L...1L,Perez2005,Hopkins2006,Karim2011ApJ...730...61K,Sobral2013MNRAS.428.1128S}. A similar behavior is found for the specific star-formation rate (sSFR, the ratio between star-formation rate (SFR) and stellar mass) at a given stellar mass \citep{Noeske2007,Dutton2010MNRAS.405.1690D,Magdis2010_IRAC,Sobral2014MNRAS.437.3516S}. However, there is still some debate about the behavior at $z > 2-3$, since there are several uncertain factors involved in the analysis, such as dust correction factors or the contribution of emission lines in the determination of stellar mass \citep{Bouwens2009ApJ...705..936B,Bouwens2012ApJ...754...83B,Stark2013ApJ...763..129S,Gonzalez2014ApJ...781...34G}. 

In agreement with the evolution of the cosmic SFR, the molecular gas content of star-forming (SF) galaxies is also higher at $z \sim 2$ than at lower redshifts, although the star-formation efficiency is claimed to not strongly depend on cosmic epoch \citep{Tacconi2010Natur.463..781T,Daddi2010ApJ...713..686D,Magdis2012ApJ...760....6M,Magdis2012ApJ...758L...9M}. The implication of this is that the increase of the cosmic SFR density with redshift is most likely driven by an increase in the molecular gas mass fraction of galaxies. A change in morphology is also found, from disk-like shapes in the local universe to clumpy, irregular, or compact morphologies at $z \sim 2$ \citep{Elmegreen2009,Malhotra2012ApJ...750L..36M,Swinbank2012ApJ...760..130S,Swinbank2012MNRAS.426..935S}. Given these rapid changes in galaxy properties it is important to ensure uniform selection of galaxy samples at different epochs.

Several selection criteria have been traditionally applied to select SF galaxies at $z \sim 2$: (1) the Ly$\alpha$ narrow-band (NB) technique, that selects Ly$\alpha$ emitters (LAEs) by sampling the redshifted Ly$\alpha$ emission line with a combination of NB and broad-band filters \cite[e.g.][]{Ouchi2008}; (2) the Lyman break technique, that selects Lyman-break galaxies (LBGs) by using colors that sample the redshifted Lyman break and includes a rest-frame UV color to rule out low-redshift interlopers \citep{Steidel2003}; (3) the $BzK$ criterion \citep{Daddi2004}, that selects SF $BzK$ ($sBzK$) galaxies with a double color selection criterion involving optical and near-IR (NIR) bands; (4) the BM/BX criteria, that select galaxies within $1 < z < 3$ with different combinations of optical broad-band filters \citep{Adelberger2004}. Other selection criteria are based on FIR/sub--mm or radio data \citep{Chapman2005,Riechers2013Natur.496..329R}, although these identify only the highest SFR systems. The NB technique has been also applied in the NIR regime with the aim to look for H$\alpha$ emitters (HAEs) at $z \sim 2$ \citep{Bunker1995MNRAS.273..513B,Moorwood2000A&A...362....9M,Kurk2004A&A...428..817K,Geach2008MNRAS.388.1473G,Hayes2010A&A...509L...5H,Lee2012PASP..124..782L,An2014ApJ...784..152A,Sobral2013MNRAS.428.1128S,Tadaki2013ApJ...778..114T}. In this way it is possible to also use H$\alpha$ to select and study SF galaxies all the way from the local Universe up to $z \sim 2$ (with H$\alpha$ moving from the optical into the $K$ band), with this being a much simpler, self-consistent and well-understood selection \citep[see][the only work so far where H$\alpha$ emission has been traced from optical to $K$ band in a single data-set/analysis]{Sobral2013MNRAS.428.1128S}.

In order to understand the nature of SF galaxies at the peak of galaxy formation and the biases and incompletenesses of each selection criterion, a comparison of the physical properties of the galaxy samples selected by the different techniques is required. This study is important for all evolutionary conclusions based on given population of galaxies, e.g. mass-metallicity relations, gas fractions, morphologies, colors, dynamics, etc \citep{Stott2013MNRAS.436.1130S,Stott2013MNRAS.430.1158S}. Furthermore, it provides a way to interpret galaxy evolution studies based on Ly$\alpha$ and Lyman-break techniques, the ones that can be applied at the highest redshifts. 

A comparison of the properties of LAEs, LBGs, and $sBzK$ galaxies has been already done \citep{Grazian2007,Ly2009,Pentericci2010,Ly2011,Haberzettl2012,Oteo2014MNRAS.439.1337O}. \cite{Haberzettl2012} found that the $BzK$ criterion is useful to select galaxies at $z \sim 2$, but the samples are biased towards massive SF galaxies and with red stellar populations. \cite{Grazian2007} report that the $sBzK$ criterion is efficient in finding SF galaxies at $z \sim 2$ but is highly contaminated by passively evolving galaxies at red $z - K_s$ colors. They also found that the Lyman-break criterion misses dusty starburst systems. \cite{Oteo2014MNRAS.439.1337O} found that most LBGs can be selected as $sBzKs$, but most of the $sBzK$ do not meet the Lyman-break criterion since the latter is biased towards blue and/or UV-bright galaxies. Furthermore, they found that $sBzK$ galaxies are similar to SF galaxies solely selected by their photometric redshift and, therefore, represent an adequate population to study the bulk of SF galaxies at $z \sim 2$. However, the $sBzK$ criterion cannot be used to carry out evolutionary studies, unlike the Lyman-break, Ly$\alpha$, or H$\alpha$ ones, due to the use of a given filter set for the galaxy selection. However, extensions to higher redshift have been proposed with other broad-band filters \citep{Guo2012}. Additionally, $sBzK$ galaxies do not represent by themselves a purely SFR-selected sample, but have a more complicated selection function. 

The main objective of this paper is studying the properties a sample of HAEs at $z \sim 2.23$ selected from the High Redshift(Z) Emission Line Survey \citep[HiZELS][]{Geach2008MNRAS.388.1473G,Sobral2009MNRAS.398...75S,Sobral2009MNRAS.398L..68S,Sobral2012MNRAS.420.1926S,Sobral2013MNRAS.428.1128S,Sobral2014MNRAS.437.3516S}. HiZELS uses a set of NB filters in NIR bands to look for emission-line galaxies up to $z \sim 9$. The study of HAEs will also allow to analyse the accuracy of the H$\alpha$ emission as a proxy of SFR and the relation between SFR and stellar mass at $z \sim 2$. Combining the H$\alpha$ observations with available, matched Ly$\alpha$ data we will also study the Ly$\alpha$ escape fraction and its relation with galaxy properties. 

This paper is organized as follows. In \S \ref{metologia} we present the data sets used, the selection of our sources, and how we analyse them. In \S \ref{comparing_things} we study the nature of HAEs and compare them with LAEs, LBGs, and $sBzK$ galaxies to place HAEs into the context of SF galaxies at the peak of cosmic star formation. In \S \ref{matched_HAEs_LAEs} we study the population of galaxies with both Ly$\alpha$ and H$\alpha$ emission and analyse the Ly$\alpha$ escape fraction at $z \sim 2$. We examine in \S \ref{sfr_ha_good} the accuracy of H$\alpha$ emission as a tracer of star formation at $z \sim 2$ and in \S \ref{sfr_mass_haes_MS} we explore the location of our galaxies in an SFR-mass plane and discuss in detail the uncertainties in the determination of the slope of the main-sequence (MS) of star formation at $z \sim 2$. Finally, the main conclusions of the work are summarized in \S \ref{conclu}.

Throughout this paper all stellar masses and SFRs reported are derived by assuming a Salpeter IMF. We assume a flat universe with $(\Omega_m, \Omega_\Lambda, h_0)=(0.3, 0.7, 0.7)$, and all magnitudes are listed in the AB system \citep{Oke1983}.

\section{Methodology}\label{metologia}

\subsection{Data sets}\label{data_sets}

Due to the availability of multi--wavelength data and NB Ly$\alpha$ and H$\alpha$ observations over an overlapping redshift range, we focus in this work on the COSMOS field \citep{Scoville2007}. In order to sample the SEDs and study the stellar populations of the galaxies analysed in this paper (see \S \ref{source_selection}), we take optical to NIR photometry from \cite{Ilbert2013A&A...556A..55I}, mid-IR IRAC and MIPS data from the S-COSMOS survey \citep{Sanders2007}, and \emph{Herschel} PACS and SPIRE data from PEP and HerMES projects, respectively \citep{Lutz2011,Oliver2012MNRAS.424.1614O}. High-quality photometric redshifts for the studied galaxies are taken from \cite{Ilbert2013A&A...556A..55I}. {\it GALEX} \citep{Zamojski2007} and {\it CHANDRA} \citep{Elvis2009} data will be also used.

\subsection{Source selection}\label{source_selection}

The main objective of this work is the analysis of the properties of NB-selected HAEs at $z \sim 2.23$. In some Sections of this paper we will use a sample of LAEs, LBGs, and $sBzK$ galaxies at $z \sim 2$ to help understand the properties of HAEs place them into the context of the SF population at $z \sim 2$. In this Section we explain how all these galaxies were selected. 

The samples of HAEs and LAEs are taken from \cite{Sobral2013MNRAS.428.1128S} and \cite{Nilsson2009}, respectively. LAEs and HAEs were selected via the NB technique, with an NB filter centered at 3963 \AA\ (129 \AA\ width) for LAEs and at 2.121 $\mu$m (210 \AA\ width) for HAEs. In addition to the NB criterion, HAE selection requires identification of the detected emission-line as H$\alpha$ at $z \sim 2.23$ rather than other line emitters at different redshifts. This selection includes a $BzK$ cut to remove low-redshift interlopers, a Lyman-break like cut to remove $z \sim 3.3$ [OIII] emitters, and inclusion of double and triple line emitters from the combination of all HiZELS NB filters. It should be pointed out that the $sBzK$ criterion applied to HAEs does not produce the loss of galaxies with unusual colors, but it is used to increase the completeness of the sample. Also, it might be possible that a small number of HAEs has been missed because of their extremely blue SEDs (similar to those for LBGs, see \S \ref{comparing_things}). However, this percentage is estimated to be very low due to the use of double and triple line detections, since blue HAEs would have very strong emission lines and low reddening, and thus be detectable in [OII] or [OIII] as well as H$\alpha$ \citep[see][for more details]{Sobral2013MNRAS.428.1128S}.

Completeness analysis indicates that LAEs are 90\% complete down to a Ly$\alpha$ flux of $f_{\rm Ly\alpha} \sim 6 \times 10^{-17} \, {\rm erg} \, {\rm cm}^{-2} \, {\rm s}^{-1}$ \citep{Nilsson2009} and HAEs are 90\% complete down to an H$\alpha$ flux of $f_{\rm H\alpha} \sim 5.6 \times 10^{-17} \, {\rm erg} \, {\rm cm}^{-2} \, {\rm s}^{-1}$ \citep{Sobral2013MNRAS.428.1128S}. We recall at this point that the Ly$\alpha$ and H$\alpha$ NB filters used in \cite{Nilsson2009} and \cite{Sobral2013MNRAS.428.1128S}, respectively, select galaxies over an overlapping redshift range. The Ly$\alpha$ filter is broader than the H$\alpha$ filter in the redshift space and, therefore, selects galaxies over a wider redshift range, with the redshift range of HAEs being fully included within the redshift range of LAEs. The great advantage is that it will be possible to study galaxies with both Ly$\alpha$ and H$\alpha$ emission even if Ly$\alpha$ has a velocity offset with respect H$\alpha$ (see \S \ref{matched_HAEs_LAEs}). 

It is important to point out that among the whole sample of 187 LAEs of \cite{Nilsson2009}, only 118 have a counterpart in the \cite{Ilbert2013A&A...556A..55I} catalog, the data set that we use for optical-to-NIR SED fits. This represents 63\% of the sample. The non-detections are mainly a consequence of the blue nature of these LAEs, that are clearly detected in $U$, $B$, $r$, or $i$ bands, but are very faint in $z$ and redder bands. The non detection in the NIR is an indication of their low stellar mass, being less massive than HAEs and other SF galaxies detected in the NIR. The significant number of LAEs without NIR counterpart indicates that the Ly$\alpha$ technique tends to select low-mass, blue galaxies. Accurate SED fits cannot be carried out for those faint LAEs due to the lack of near-IR information that is essential for age, stellar mass, and redshift estimations. Stacking might be an alternative, but it has been reported that stacking in LAEs does not provide reliable estimations of the median properties of the population \citep{Vargas2014ApJ...783...26V}. Therefore, we have decided not to include these LAEs in the analysis. This might bias our results since we only include in the final sample the most massive LAEs selected through the NB technique in \cite{Nilsson2009}. We will indicate the implications of this in the relevant sections of the paper. 

Regarding HAEs, most of them have a counterpart in the \cite{Ilbert2013A&A...556A..55I} catalog, with only 8\% being undetected. The non detections are mainly due to the faintness of that small population of HAEs in optical bands. At the same time this is an indication that the HAE selection is also able to identify very dusty sources that might be missed in optical-based studies. We do not include the previous 8\% of faint HAEs in our analysis since their UV continuum cannot be well constrained. They represent a very low percentage of the total sample and, therefore, their exclusion is not expected to affect significantly the conclusions presented in this work. However, it should be noted that due to their red colors (e.g. high dust extinction) they might be a significant proportion of dust-extinguished HAEs.

LBGs and $BzK$ galaxies are taken from \cite{Oteo2014MNRAS.439.1337O}. LBGs were selected with the classical dropout technique, where the NUV and $U$ bands were used to sample the Lyman break at $z \sim 2$ and a $U$-$V$ color was used to avoid contamination from low-redshift interlopers. Furthermore, non detection in the {\it GALEX} FUV band was imposed. The $BzK$ galaxies were selected using the criterion of \cite{Daddi2004}, that picks up both SF galaxies ($sBzK$ galaxies) and quiescent galaxies ($pBzK$ galaxies). Since we are interested in galaxies dominated by star formation, we only consider $sBzK$ galaxies for most of the analysis presented in this work, although $pBzK$ galaxies will be used for a comparison in some discussions. While LAEs and HAEs have a narrow redshift distribution due to their selection with NB filters, LBGs and $sBzK$ have redshifts spanning typically $1.5 < z < 2.5$ \citep{Daddi2004,Oteo2014MNRAS.439.1337O}. Therefore, in order to carry out a fairer comparison with LAEs and HAEs at $z \sim 2.23$, we additionally limit the photometric redshift of LBGs and $sBzK$ galaxies to $2.0 < z_{\rm phot} < 2.5$. This redshift range has been selected to account for the uncertainties of photometric redshift determinations at $z \sim 2.25$ \citep{Ilbert2013A&A...556A..55I}. To avoid a possible presence of any low-redshift interlopers in the HAE and LAE samples, we also limit their photometric redshifts to the same range: $2.0 < z_{\rm phot} < 2.5$. Again, this range is chosen to account for the $z_{\rm phot}$ uncertainties.

We clean all samples from AGN contamination by removing all sources with X-ray {\it CHANDRA} detections \citep{Elvis2009}. It should be pointed out that the percentage of X-ray detections in our galaxies is low, less than 5\% in all the four samples studied. Additionally, we use {\it GALEX} photometry \citep{Zamojski2007} to clean our samples from low-redshift interlopers. At $z \sim 2$ the Lyman break is redshifted out of the UV regime and, therefore, our galaxies cannot be detected in {\it GALEX} bands.

After all these considerations, we end up with a sample of 373 HAEs, 69 LAEs, 3751 LBGs, and 13194 $sBzK$ galaxies. We note that out of all the samples, HAEs are the ones drawn from the smallest volume, followed by LAEs, LBGs, and $sBzK$ galaxies. Thus, the number of galaxies in each sample is largely driven by the different volumes. The number density of HAEs is $4.8 \times 10^{-4} \, {\rm Mpc}^{-3}$ down to $\log{\left( L_{\rm H\alpha} \right)} > 42.0$ and the number density of LAEs is $1.6 \times 10^{-4} \, {\rm Mpc}^{-3}$ down to $\log{\left( L_{\rm Ly\alpha} \right)} > 42.3$. The number density of LAEs obtained here is smaller than the value reported in \cite{Nilsson2009}, mostly due to our inclusion of the criterion to clean the sample from lower-redshift interlopers and because we only include in the sample galaxies detected in the NIR. Furthermore, the value reported in \cite{Nilsson2009} was calculated over 28\% of the area covered by their observations, which in turn represents 2\% of the entire area of the COSMOS field. This makes their calculation very uncertain due to the influence of cosmic variance. The number density of LBGs is $3.1 \times 10^{-4} \, {\rm Mpc}^{-3}$, and the number density of $sBzK$ galaxies is the highest in our samples, $1.1 \times 10^{-3} \, {\rm Mpc}^{-3}$, due to the high number of sources selected. Also note that LBGs and $sBzK$ have different, more uncertain SFR limits.

\subsection{Analysis}\label{SED_fitting_procedure}

In the rest-frame UV to NIR regime we analysed the nature of our selected galaxies via the traditional SED-fitting technique with \cite{Bruzual2003} (hereafter BC03) templates. To this end we used \emph{LePhare} \citep{Arnouts1999MNRAS.310..540A,Ilbert2006A&A...457..841I}. We included the emission lines in the stellar population templates \citep{Schaerer2009A&A...502..423S,deBarros2014A&A...563A..81D} since we are working with SF galaxies and, in fact, LAEs and HAEs are selected through their emission lines. The strength of the Ly$\alpha$ emission is more uncertain than other emission lines due to its resonant nature. Therefore, we do not include $U$-band information in the fits. This does not significantly change the values of the SED-derived properties, since the UV continuum is well sampled with the other filters. In this way, the filters used in the fits are: Subaru $B_J$, $V_J$, $r^+$, $i^+$, $z^+$, and VISTA $Y$, $J$, $H$, and $K_s$ \citep{Ilbert2009ApJ...690.1236I,McCracken2012A&A...544A.156M}. IRAC data for the IRAC-detected galaxies are also included \citep{Sanders2007}. 

The BC03 templates used in this work were built by assuming an exponentially declining SFH and a fixed metallicity $Z = 0.2 Z_\odot$. We considered time scale $\tau_{\rm SFH}$ values of 0.1, 0.2, 1.0, 2.0, 3.0, and 5.0 Gyr. We chose a fixed value for metallicity because this parameter tends to suffer from large uncertainties \citep[see for example][]{deBarros2014A&A...563A..81D}. For age we considered values ranging from 10 Myr to 3.4 Gyr, the age of the Universe at the median redshift of our galaxies. Ages values were taken in steps of 10 Myr from 10 to 100 Myr, in steps of 20 Myr from 100 to 200 Myr, in steps of 50 Myr from 200 to 500 Myr, in steps of 100 Myr from 500 Myr to 1 Gyr, and in steps of 0.2 Gyr from 1.0 to 3.4 Gyr. Dust attenuation was included in the templates via the \cite{Calzetti2000} law and parameterized through the color excess in the stellar continuum, $E_s(B-V)$, for which values ranging from 0 to 0.7 in steps of 0.05 were considered. 

\begin{figure*}
\centering
\includegraphics[width=0.95\textwidth]{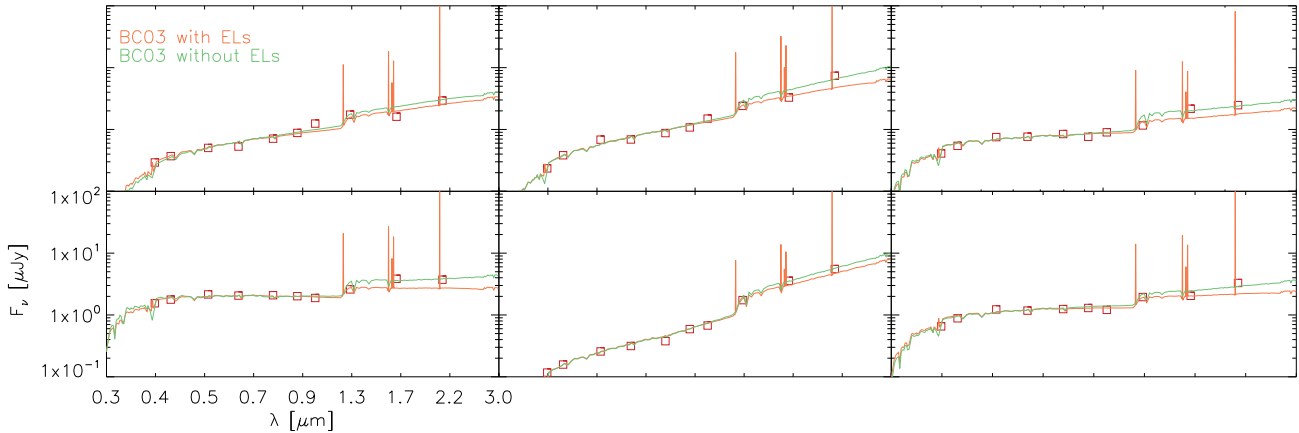}
\caption{Examples of SED fit results for 6 randomly-selected HAEs from our sample. These fits are representative of the whole sample of HAEs studied in this work. The best-fitted BC03 templates including emission lines are shown in orange, and best-fitted templates with no emission lines are in green. Red open squares are the observed photometric data. Due to the uncertain contribution of the Ly$\alpha$ emission, the $U$ band information has not been included in the fits. Our HAEs (and also LAEs) are selected for having strong H$\alpha$ (and Ly$\alpha$) emission. Therefore, it is expected that emission lines affect their observed fluxes. It can be seen in this figure that the best-fitted templates including emission lines have fainter rest-frame optical continua and, therefore, represent less massive galaxies. If templates with no emission lines had been used, the stellar masses would have been overestimated. The UV continuum (and hence ${\rm SFR_{UV}}$) of the best-fitted templates does not change significantly when emission lines are included.
              }
\label{optical_SEDs_HAEs}
\end{figure*}

Once the templates are fitted, the rest-frame UV luminosity for each galaxy was obtained from its normalized best-fitted template and converted to ${\rm SFR_{UV}}$ via the \cite{Kennicutt1998} calibration. Note that ${\rm SFR_{UV}}$ is not corrected for dust attenuation. The UV continuum slope was obtained by fitting a power-law function to the UV continuum of each best-fitted template. Figure \ref{optical_SEDs_HAEs} shows the SED fit results for six HAEs randomly selected from our sample. For a reference, the best-fitted templates with no inclusion of emission lines (fitted with \emph{LePhare} as well) are also included. It can be seen that emission lines have a clear effect on the observed fluxes and the best-fitted rest-frame optical continuum emission. This is because the strongest rest-frame optical emission lines are sampled with some broad-band filters used in the fits: H$\alpha$ is within the $K_s$ band, [OIII]5007 within the $H$ band, and [OII]3727 within the $J$ band. While the rest-frame UV SEDs are similar in all the cases, fainter rest-frame optical continua are obtained when including the effect of emission lines. This translates into lower stellar masses and younger ages.

\begin{figure*}
\centering
\includegraphics[width=0.33\textwidth]{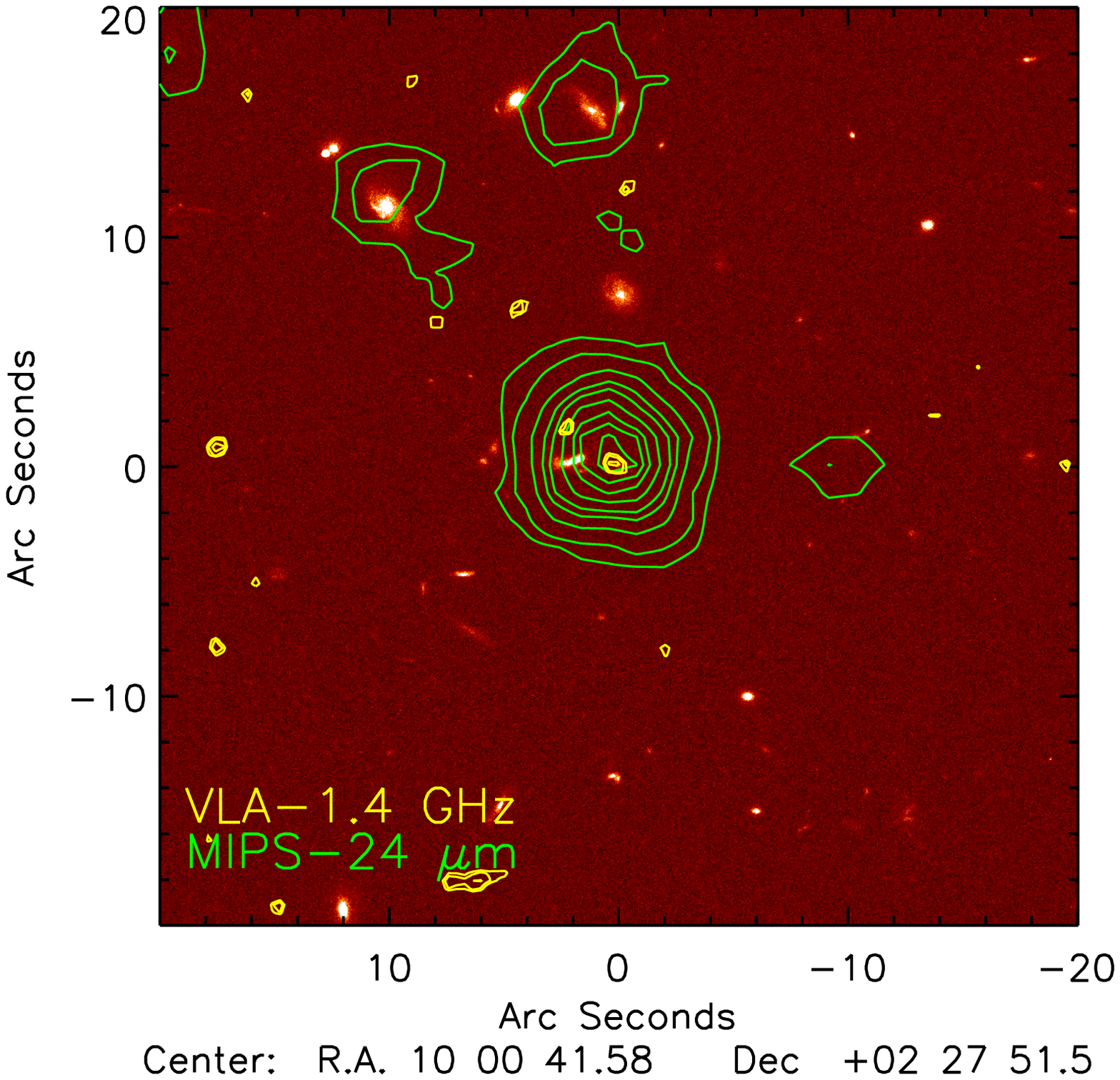}
\includegraphics[width=0.33\textwidth]{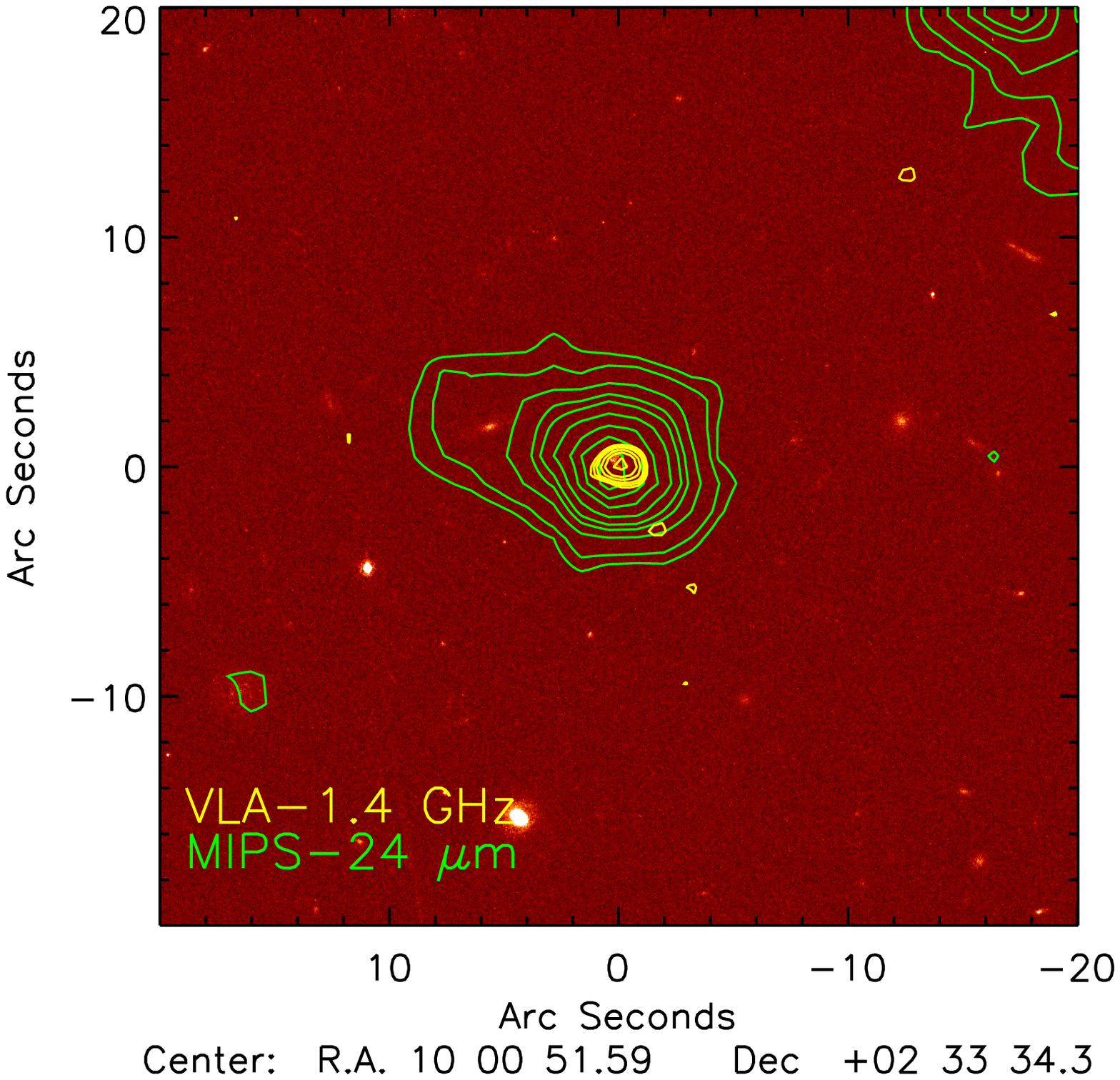}
\includegraphics[width=0.33\textwidth]{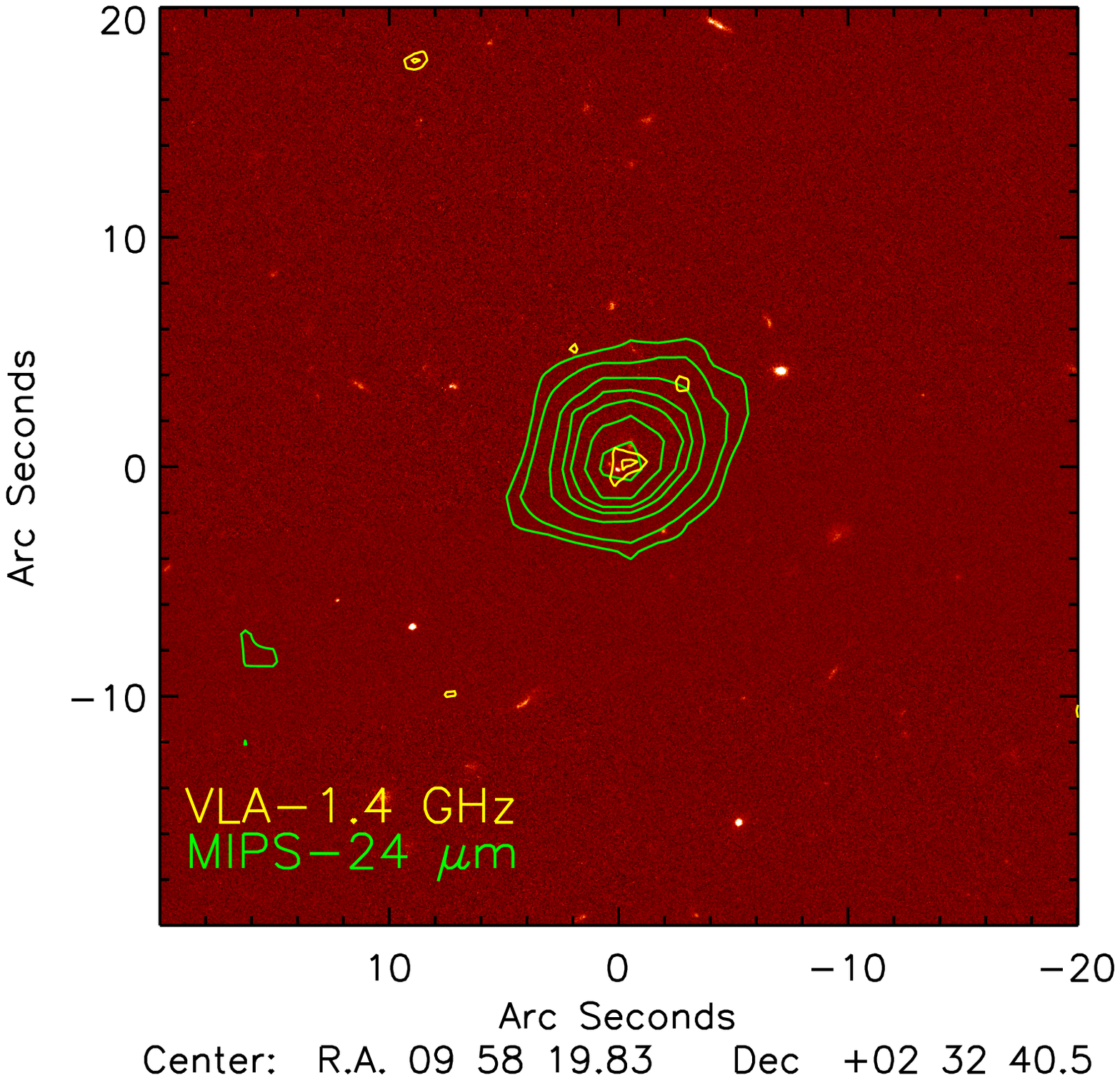}
\caption{Analysis of source confusion and contamination at FIR wavelengths from nearby sources in the three SPIRE-500 $\mu$m-detected HAEs. The MIPS-24 $\mu$m (green), and VLA (yellow) contours are over-plotted in the ACS I-band images of the galaxies. HAEs are located in the center of each image. Images are 40$''$ on each side, slightly larger than the SPIRE-500 $\mu$m beam, the \emph{Herschel} band with the largest PSF. The VLA and MIPS detections, that have better spatial resolution than PACS and SPIRE, indicate that there is no significant contribution of possible nearby FIR-bright sources that might contaminate the HAE \emph{Herschel} fluxes. The PSF of PACS bands is much smaller than the SPIRE PSFs, so contamination is even more unlikely.
              }
\label{source_confusion_contours}
\end{figure*}

Among the whole sample of SF HAEs, only nine are individually detected in any of the \emph{Herschel} bands. This represents a detection rate of 3\%. Interestingly, despite being very low, the percentage of \emph{Herschel} detections is higher for HAEs than for LBGs ($\sim$ 0.7 \%) or $sBzK$ galaxies ($\sim$ 0.5 \%). This indicates that the H$\alpha$ selection can recover not only relatively dust-free, but also highly dust obscured sources. However, the comparison between the number of \emph{Herschel} detections and physical properties is challenging since at $z \sim 2$ \emph{Herschel} only selects the most extreme galaxies rather than normal SF galaxies. Source confusion is also a major problem when analysing the FIR emission of UV, optical, or NIR-selected galaxies. We have attempted to identify source confusion by analysing the optical ACS images of the galaxies along with the MIPS-24 $\mu$m and VLA contours. As an example, see results in Figure \ref{source_confusion_contours} for the three SPIRE-500 $\mu$m-detected HAEs. The size of the images are 40$''$ on each side, slightly larger than the SPIRE-500 $\mu$m beam, the \emph{Herschel} band with the largest PSF. In the three cases, the location of the MIPS detection is coincident with a radio emission, indicating that there is no significant contribution of nearby FIR-bright sources that might contaminate the fluxes in SPIRE bands. The contamination is even more unlikely in PACS bands, since their PSF is 2--4 times smaller than SPIRE beams. As a sanity check for SPIRE-detected sources, we have re-done the FIR SED fits including only their less-likely contaminated PACS photometry. The values obtained for the total IR luminosity, and hence for ${\rm SFR_{IR}}$, are in agreement with those including SPIRE data within the uncertainties ($\sim 0.1-0.2$ dex). It should be pointed out that there are two LAEs individually detected in \emph{Herschel} but they are also detected in X-ray and, consequently, have a likely AGN nature and are not considered in this work (see \S \ref{source_selection} and \citealt{Bongiovanni2010}). 

\begin{figure*}
\centering
\includegraphics[width=0.33\textwidth]{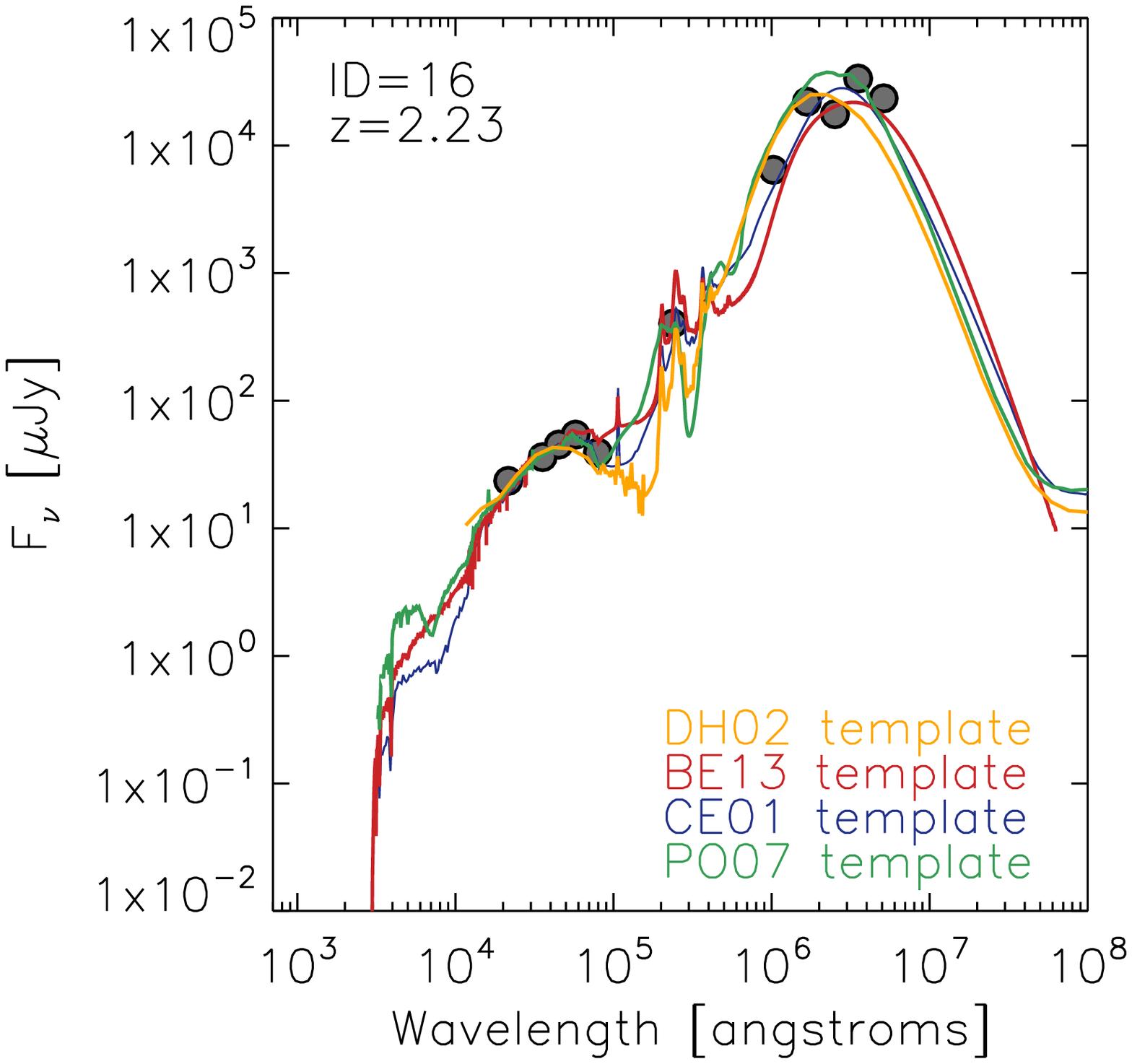}
\includegraphics[width=0.33\textwidth]{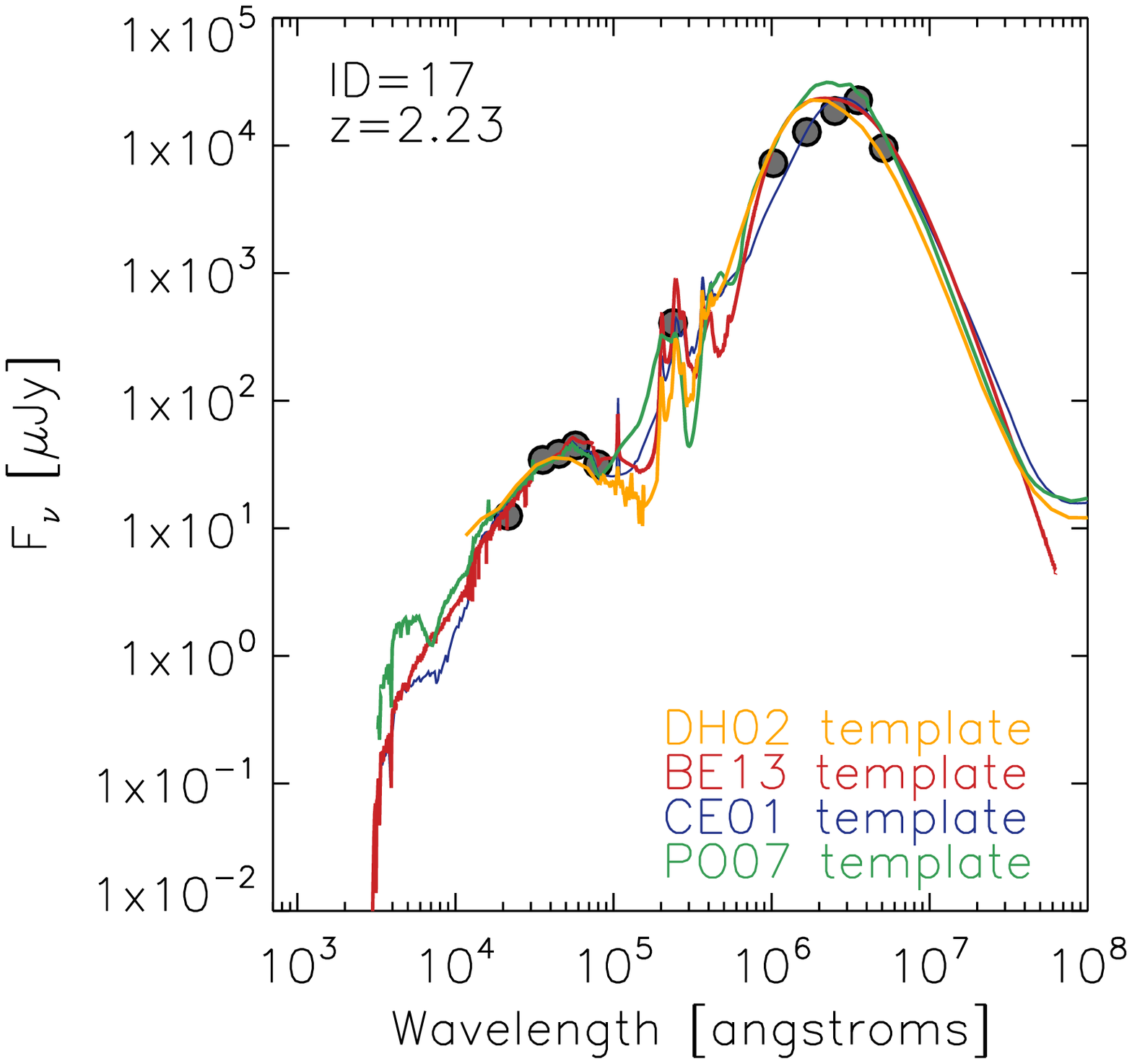}
\includegraphics[width=0.33\textwidth]{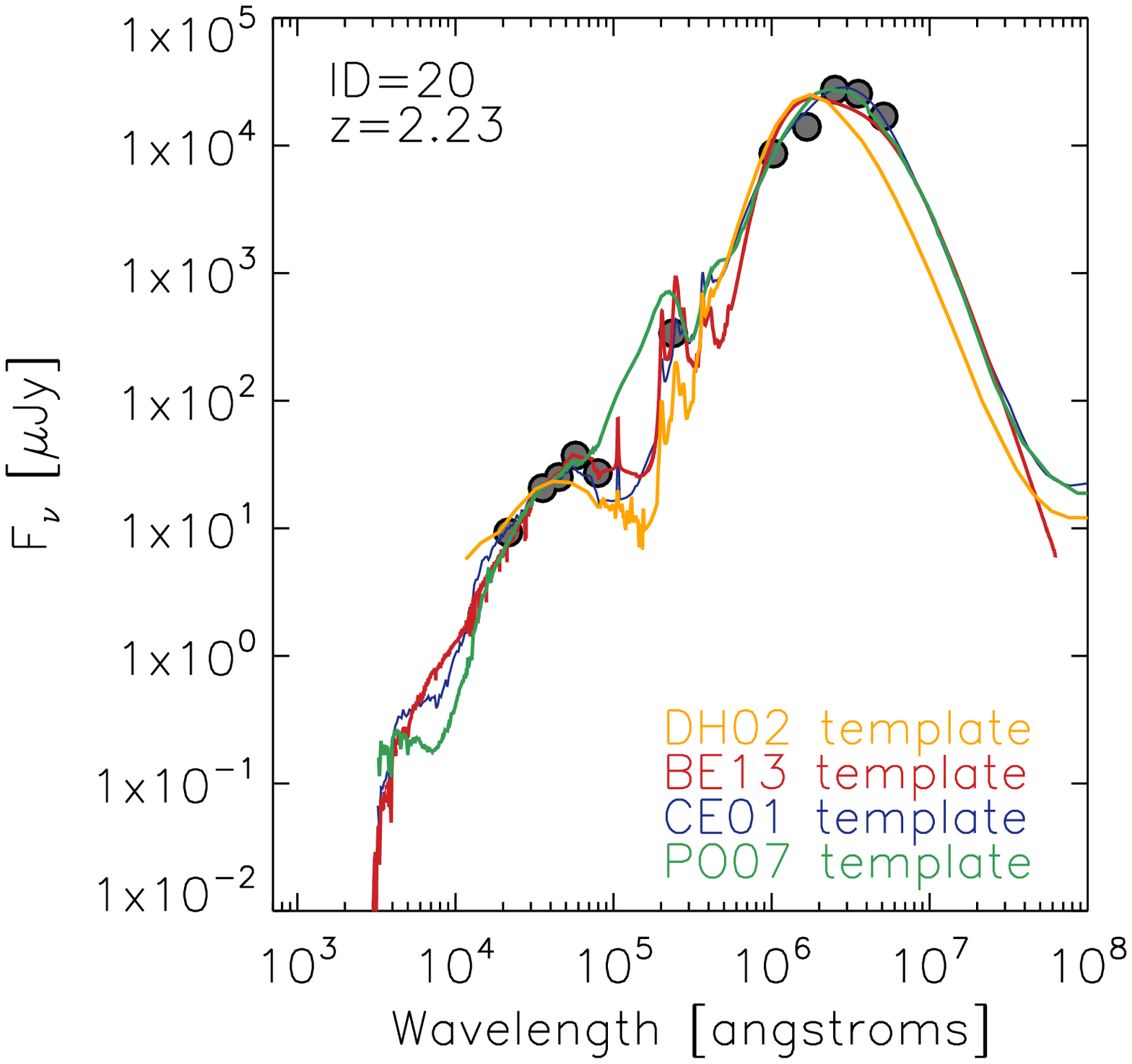}
\caption{IR SEDs of the three SPIRE-500 $\mu$m-detected HAEs as an illustration of the SED-fitting results for the \emph{Herschel}-selected galaxies studied in this work. We plot the observed photometry in $K_s$, the four IRAC bands, MIPS-24 $\mu$m, PACS, and SPIRE. The best fitted \citet{Chary2001}, \citet{Polletta2007}, \citet{Dale2002}, and \citet{Berta2013A&A...551A.100B} templates are also shown, with the color code indicated in the bottom-right legend. In the SED fits, a redshift of $z = 2.23$ has been assumed. The data provide a good sampling of the dust emission peak in HAEs at $z \sim 2.23$. Therefore, the integration of the best-fitted templates between 8 and 1000 $\mu$m provides an accurate determination of their total IR luminosities and, consequently, dust-corrected SFR. No significant difference is found for $L_{\rm IR}$ when using different templates. We choose to report results obtained with \citet{Chary2001} templates, as in many previous works.              }
\label{IR_SED_HAEs_SPIRE}
\end{figure*}

The IR SEDs of the FIR-detected galaxies are fitted with \cite{Chary2001} (CE01), \cite{Dale2002}, \cite{Polletta2007}, and \cite{Berta2013A&A...551A.100B} templates. As an example, we show in Figure \ref{IR_SED_HAEs_SPIRE} the IR SEDs of the three HAEs detected in SPIRE-500 $\mu$m. The presence of the rest-frame 1.6 $\mu$m stellar bump (sampled with the IRAC bands at the redshift of our galaxies) is clear in the three galaxies, indicating that their SEDs are dominated by star formation. All the different templates fit well the IR SEDs of the galaxies, with similar values of $\chi^2_r$. We choose to report the results obtained with CE01 templates, as in many previous works in the literature. The best-fitted CE01 templates are integrated between rest-frame 8 and 1000 $\mu$m to derive total IR luminosities, $L_{\rm IR}$. These are then converted into SFR by using the \cite{Kennicutt1998} calibration. The total SFR is then obtained by assuming that all the light absorbed by dust in the UV is reemitted in the FIR: ${\rm SFR_{total} = SFR_{UV} + SFR_{IR}}$.

Since most galaxies are not individually detected in the FIR, we also performed stacking analysis in \emph{Herschel} bands to study the FIR emission of \emph{Herschel}-undetected galaxies \citep[see for example][for stacking analysis in HAEs at $z \sim 1.47$]{Ibar2013MNRAS.434.3218I}. One single band near the peak of the dust SED is enough to estimate the total IR luminosity. We focused on the PACS-160 $\mu$m band for stacking due to its relatively small beam and employ the residual maps, as in many previous works. We stacked by using the publicly available IAS Stacking Library \citep{Bethermin2010A&A...512A..78B} and uncertainties in the stacked fluxes were derived by using bootstraps. For HAEs, we stacked in different bins of stellar mass (from $\log{\left( M_* /M_\odot \right)}$ = 9.5 to 11 in bins of 0.5 dex) and dust-corrected H$\alpha$-derived SFR (${\rm SFR_{H\alpha}}$) (from $\log{\left({\rm SFR_{\rm H\alpha}}\right [\rm{M_\odot \, yr^{-1}}] )} = 1.0$ to 2.0 in bins of 0.5). Those bins cover the whole range of values for those parameters. For $sBzKs$, where the number of sources is high and allows stacking over more stellar mass bins, we stacked from $\log{\left( M_* /M_\odot \right)}$ = 9.8 to 11.6 in bins of 0.2 dex.

We only detected stacked fluxes for HAEs in the bin corresponding to the highest stellar mass ($10.5 \leq \log{\left(M_*/M_{\odot}\right)} < 11$, where 66 galaxies are included and the stacked flux is $f_{\rm 160 \mu m} = 1.4 \pm 0.3$ mJy, corresponding to $\log{\left( L_{\rm IR}/L_\odot \right)} = 11.8 \pm 0.1$) and highest ${\rm SFR_{H \alpha}}$ ($1.5 \leq \log{\rm \left(SFR_{\rm H\alpha}\right)} < 2.0$, where 166 sources are included and the stacked flux is $f_{\rm 160 \mu m} = 1.0 \pm 0.2$ mJy, corresponding to  $\log{\left( L_{\rm IR}/L_\odot \right)} = 11.3 \pm 0.1$). This is consistent with more massive SF galaxies being more affected by dust, in agreement with \cite{Garn2010MNRAS.409..421G}, \cite{Sobral2012MNRAS.420.1926S} or \cite{Ibar2013MNRAS.434.3218I}, although it could be also due to a pure scaling of the SED. We also stacked LAEs as a function of stellar mass (with the same bins as for HAEs), but no stacked detections are recovered probably due to the low number of sources producing poor statistic and also to the less dusty nature of LAEs (see also \S \ref{comparing_things}). In $sBzK$ galaxies we recover detections for $9.8 < \log{\left( M_* /M_\odot \right)} < 11.0$ with more than 1000 galaxies in each bin. The recovered stacked fluxes are summarized in Table \ref{hola_tabla}. The stacked PACS-160 $\mu$m fluxes were converted into $L_{\rm IR}$ by using single band extrapolations with CE01 templates and errors were obtained from the flux uncertainties. SFR$_{\rm IR}$ were obtained with the \cite{Kennicutt1998} calibration.

\begin{table}\label{hola_tabla}
\begin{center}
\caption{Stacked PACS-160 $\mu$m fluxes and their associated total IR luminosities for our sample of $sBzK$ galaxies at $2.0 < z_{\rm phot}$ < 2.5. }
\begin{tabular}{c c c }
\hline
Stellar mass range & Stacked $f_{160 \mu m}$ & $\log{\left( L_{\rm IR}/L_\odot \right)}$\\ 
\hline
\hline

$9.8 \leq   \log{\left( M_* / M_\odot \right)} < 10.0$ &     $0.53 \pm 0.05$ mJy 	&    	 $11.4 \pm 0.1$ \\
$10.0 \leq \log{\left( M_* / M_\odot \right)} < 10.2$ &     $0.80 \pm 0.06$ mJy  &     $11.5 \pm 0.1$ \\
$10.2 \leq \log{\left( M_* / M_\odot \right)} < 10.4$ &     $1.00 \pm 0.09$ mJy  &     $11.6 \pm 0.1$ \\
$10.4 \leq \log{\left( M_* / M_\odot \right)} < 10.6$ &     $1.20 \pm 0.12$ mJy  &     $11.7 \pm 0.1$ \\
$10.6 \leq \log{\left( M_* / M_\odot \right)} < 10.8$ &     $1.23 \pm 0.12$ mJy &     $11.7 \pm 0.1$ \\
$10.8 \leq \log{\left( M_* / M_\odot \right)} < 11.0$ &     $1.40 \pm 0.17$ mJy  &     $11.8 \pm 0.1$ \\
\hline
\end{tabular}
\end{center}
\end{table}

\section{The nature of H$\alpha$ emitters at $z \sim 2$}\label{comparing_things}

\begin{figure*}
\centering
\includegraphics[width=\textwidth]{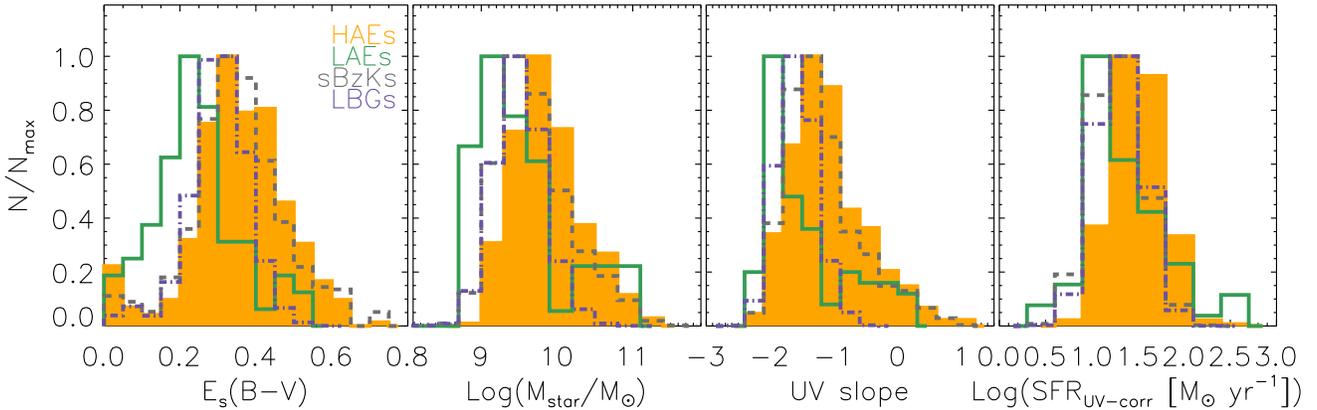}
\caption{From left to right: Dust attenuation, stellar mass, UV continuum slope, and dust-corrected total SFR of our selected HAEs (orange filled histograms). We include the distributions obtained for LAEs, LBGs, and $sBzK$ galaxies. These properties were obtained by fitting BC03 templates to the observed multi-wavelength photometry (from $B$ to IRAC bands, when available) of each galaxy. The BC03 templates were built by assuming an exponentially declining SFH and a fixed metallicity of $Z = 0.2 Z_\odot$. Emission lines were also included in the templates. Histograms have been normalized to their maxima in order to clarify the representations. The distributions indicate that HAEs and $sBzK$ galaxies have similar stellar populations, both in the median values and in the range covered. HAEs have a very-well defined selection criterion and, therefore, represent an excellent sample for studies of star formation at $z \sim 2$. LAEs are significantly biased towards blue, less massive, and dust-poor galaxies.
           }
\label{properties_SEDs_comparison}
\end{figure*}

Figure \ref{properties_SEDs_comparison} shows the SED-derived dust attenuation, stellar mass, UV continuum slope, and dust-corrected SFR of our HAEs at $z \sim 2.23$. It can be seen that HAEs can be either dust-poor (and have blue UV slopes) or dusty (and have red UV slopes) and have a wide range of stellar masses and dust-corrected SFR. This already indicates that our sample of SF HAEs, selected down to a fixed dust-uncorrected SFR, is not significantly biased towards any SED-derived property. The distributions of SED-derived properties of HAEs resemble those for $sBzK$ galaxies, one of the classical populations traditionally used to study galaxy properties at $z \sim 2$. However, HAEs seem to have slightly higher stellar masses and SFR on average, probably due to their selection based on SFR. The distributions for HAEs contrast to those found for LAEs, which have much lower dust attenuation and stellar mass and much bluer UV continuum slopes on average. This indicates that Ly$\alpha$ and H$\alpha$ samples are formed, on average, by galaxies with different stellar populations (although the values obtained for LAEs are within the distributions found for HAEs). This suggests that the selection based on Ly$\alpha$ is much more biased than the selection in H$\alpha$ and that using Ly$\alpha$ to select high-redshift galaxies might lead to the loss of a significant population at a given redshift (mostly the reddest and most massive galaxies). Figure \ref{properties_SEDs_comparison} shows that the sample of LBGs is also slightly biased towards galaxies with lower dust attenuation and bluer UV continuum slope, although the effect is not as strong as it is for LAEs.

\begin{figure}
\centering
\includegraphics[width=0.45\textwidth]{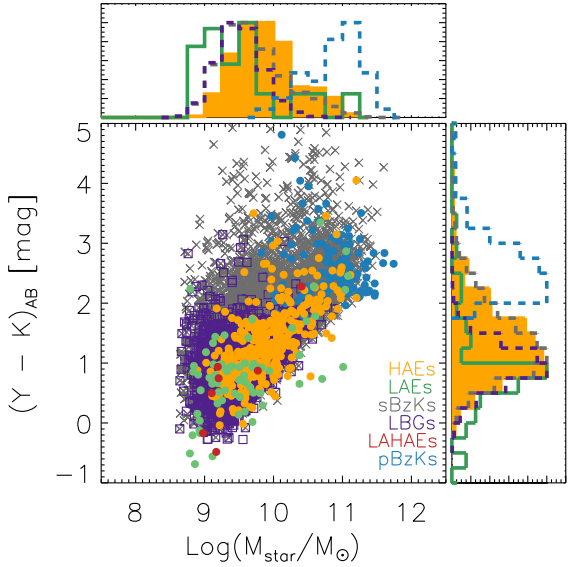}
\caption{Relation between the $ Y - K$ color and stellar mass for our sample of HAEs at $z \sim 2.25$. We also represent the location of LAEs, LBGs, and $sBzK$ galaxies selected in the same field. At the redshift of our galaxies, $Y-K$ matches the rest-frame $u - r$ color used to study the color distribution and define the blue cloud and red sequence in local galaxies \citep{Strateva2001}. For a reference we also plot the location of the $pBzK$ quiescent population \citep{Daddi2004} and LAHAEs (see \S \ref{matched_HAEs_LAEs}). We also represent the distributions of $Y-K$ color and stellar masses for each kind of galaxy but not for LAHAEs due to their low number. It can be clearly seen that HAEs are well distributed across a wide range of colors and stellar masses as are $sBzK$ galaxies. However, LAEs and LBGs are the galaxies with the bluest colors and lowest masses, indicating the biased nature of those selections. LBGs and $sBzK$ galaxies are distributed over a larger area due to their wider photometric redshift distributions. Most LAEs are blue and less massive, but there is a small population of LAEs with red colors. Although the number of such red LAEs is not high, it indicates that Ly$\alpha$ emission can also escape from dusty, red systems, as previously reported from individual detections in FIR wavelengths.
           }
\label{color_masa}
\end{figure}

Figure \ref{color_masa} shows the relation between color and stellar mass for our HAEs. We choose the $Y-K_s$ color because it matches at $z \sim 2$ with the rest-frame $u-r$ color traditionally used to define the blue cloud and the red sequence in the local Universe \citep[see for example][]{Strateva2001}. Defining a clear difference between the blue cloud and red sequence at $z \sim 2$ is challenging due to the low number of passive galaxies populating the red sequence. However, we consider here that $pBzK$ galaxies represent the prototype of passive galaxies populating the red sequence at $z \sim 2$. These galaxies are also represented in Figure \ref{color_masa}. HAEs cover a wide range of colors and stellar masses and, thus, they represent a diverse population with a range of properties. H$\alpha$ selection only misses a small population of the bluest and least massive galaxies at $z \sim 2.25$. Some SF HAEs have colors even similar to $pBzK$ galaxies, as happens for obscured galaxies with intense, obscured star formation \citep{Oteo2013MNRAS.435..158O,Oteo2014MNRAS.439.1337O}. 


Interestingly, although most galaxies with Ly$\alpha$ emission have low dust attenuation and blue colors, there is a population of 12 red LAEs with $\log{\left( M_* / M_\odot \right) > 10.25}$. Their stellar masses are as high as the most massive HAEs. All 12 red LAEs are detected in IRAC and there is no indication of power-law like MIR SED, so these red LAEs are SF galaxies rather than AGNs (note that AGN contamination in all our samples was avoided by discarding galaxies with X-ray detection). This population represents a low fraction of the whole LAE sample, but it indicates that Ly$\alpha$ emission can also escape from dusty, massive, and red galaxies, as previously shown via optical colors \citep{Stiavelli2001ApJ...561L..37S} and submm or FIR detections at different redshifts \citep{Chapman2005,Oteo2012A&A...541A..65O,Oteo2012ApJ...751..139O,Casey2012_z2,Sandberg2015arXiv150106017S}. Ly$\alpha$ surveys over larger areas would be needed to increase the samples of red and massive LAEs and study in detail how and why Ly$\alpha$ can escape from dusty galaxies.

\begin{figure}
\centering
\includegraphics[width=0.45\textwidth]{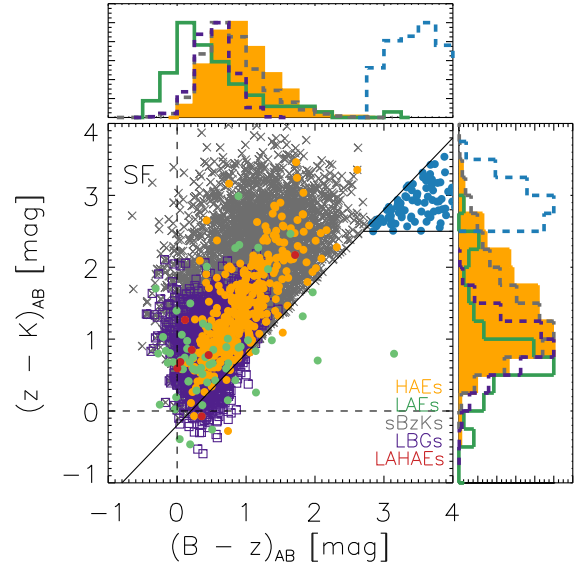}
\caption{Location of our HAEs at $z \sim 2.25$ in the color-color diagram employed to look for $BzK$ galaxies at $z \sim 2$ \citep{Daddi2007}. We also represent the location of LAEs, LBGs and LAHAEs (see \S \ref{matched_HAEs_LAEs}). The distributions of $B - z$ and $z - K$ colors are also represented for each kind of galaxy but not for LAHAEs due to the low number of galaxies in that sample. It can be seen that most LAEs, LBGs, and HAEs can also be selected through the $sBzK$ criterion. There is a small population of LBGs (and less numerous sample of LAEs) that are missed through the $sBzK$ criterion, indicating that it misses the bluest and youngest galaxies at $z \sim 2$.
              }
\label{color_bzk_figure}
\end{figure}

It is also important to examine the overlap between HAEs and other populations of SF galaxies at $z \sim 2.25$. In other words: what is the fraction of galaxies of a given type which could have been selected/missed through any of the other criteria? This study has been already done for LBGs, UV-selected, and $sBzK$ galaxies in \cite{Oteo2014MNRAS.439.1337O}. In that work, it was concluded that most LBGs can be also selected through the $sBzK$ criterion and only 25\% of $sBzK$ galaxies would have been selected as LBGs (mainly due to the bias of the Lyman-break selection towards UV-bright galaxies). Consequently, $sBzK$ galaxies are a better representation of the bulk of SF galaxies over $1.5 < z < 2.5$ than LBGs. 

Figure \ref{color_bzk_figure} represents the classical color-color diagram employed to select high-redshift $BzK$ galaxies, both SF and passively evolving \citep{Daddi2004}. Studying the location of our selected HAEs in such a diagram gives information about the overlap between those populations. Most HAEs can be also selected as $sBzK$ galaxies. As commented in \S \ref{source_selection}, this was expected for HAEs since a $BzK$ selection was applied in \cite{Sobral2013MNRAS.428.1128S} to the NB-selected HAEs to avoid contamination from low-redshift interlopers. HAEs cover the same range of colors than $sBzK$ galaxies confirming that they are not strongly biased either towards dust obscured or dust-free objects. However, as suggested by Figure \ref{color_masa}, the H$\alpha$ selection might miss the bluest galaxies (now in terms of their $z - K$ color) at $z \sim 2.25$, as it happens to $sBzK$ galaxies. Actually, there is a sub-population of 'blue' LBGs with $BzK = (z-K)_{\rm AB} - (B-z)_{\rm AB} \leq -0.2$ and $(B-z)_{\rm AB} < 1$ that cannot be selected as $sBzK$ galaxies. This is mainly due to their blue $z - K$ color for their $B-z$ color. At $2 < z < 2.5$, the $z - K$ color samples the Balmer and 4000 \AA\ breaks and, therefore, is a proxy of the age of the galaxies: younger galaxies should have bluer $z - K$ colors. This is further confirmed by the shape of their optical SEDs and ages, since 80\% of the blue LBGs are younger than 100 Myr (always under the assumption of exponentially declining SFH and $0.2 Z_\odot$ metallicity and with the caution of the degeneracies between age, SFH, dust attenuation, and metallicity) with nearly flat SEDs (when flux densities are expressed per frequency units). Finally, the colors of LAEs indicate that they are more biased towards bluer galaxies, in agreement with the results obtained from SED fits and Figure \ref{color_masa}. 

A significant percentage of HAEs cannot be selected as LBGs due to their faint rest-frame UV continuum. This indicates that a significant population of galaxies with intense H$\alpha$ emission, and thus actively forming stars, would be missed if only the drop-out criterion had been applied. This, along with the relatively high percentage of $sBzK$ galaxies that could be missed in the drop-out criterion used in this work, suggests that the Lyman-break technique employed in this work is biased towards UV-bright galaxies and, consequently, would not be adequate for studying the properties of a complete census of galaxies at the peak of star formation. This result is important for evolutionary studies \citep{Stott2013MNRAS.436.1130S,Stott2013MNRAS.430.1158S,Sobral2014MNRAS.437.3516S}. 

\begin{figure}
\centering
\includegraphics[width=0.49\textwidth]{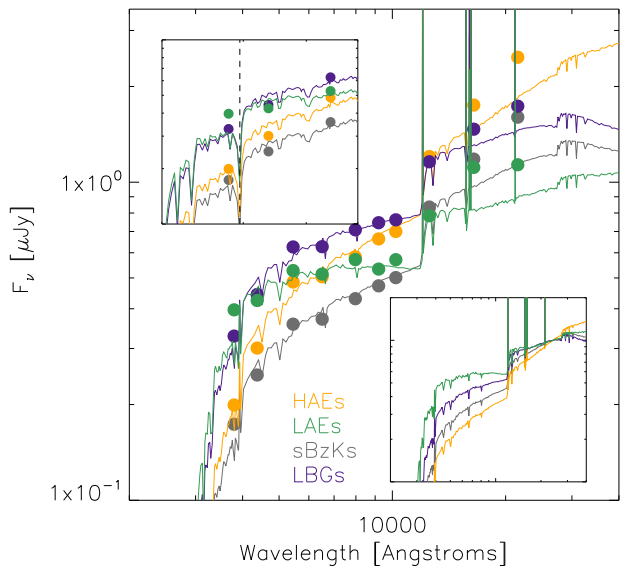}
\caption{Median SEDs of the HAEs, LAEs, LBGs and $sBzK$ galaxies studied in this work. Bottom-right panel shows the median SEDs renormalized to unity at 2.5 $\mu$m. In order to explore the EW of the Ly$\alpha$ emission in each sample we show the region around Ly$\alpha$ in the upper-left panel. In this panel, the location of Ly$\alpha$ emission is indicated with a vertical dashed line. This Figure shows that LBGs have a bright rest-frame UV continuum, whereas $sBzK$ galaxies are the most attenuated on average. Except for the difference in the normalization factors, the median SED of HAEs and $sBzK$ galaxies are similar, in agreement with the similarity in their stellar populations. The upper-left panel indicates that the Ly$\alpha$ emission boosts the $U$-flux only in LAEs, indicating that LBGs, HAEs, and $sBzK$ galaxies do not have strong Ly$\alpha$ with high equivalent width, on average.
      }
\label{figure_median_SEDs}
\end{figure}

To further analyse the properties of our HAEs we study now their median rest-frame UV-to-optical SED and compare it to the median SED of LAEs, LBGs, and $sBzK$ galaxies. The median SEDs were built from the median fluxes in each photometric band. Then, following the same procedure presented in \S \ref{SED_fitting_procedure} we fitted their observed fluxes with BC03 templates including emission lines. In these fits we did not include the $U$ band photometry because it might be potentially contaminated by the uncertain strength of the Ly$\alpha$ emission, not included in the templates. The results are shown in Figure \ref{figure_median_SEDs}. Apart from the normalization factor, the shape of the median SED of HAEs and $sBzK$ galaxies is similar, supporting previous results derived with SED-derived properties and observe colors. The SED of HAEs constrast with the median SED of LAEs, which is much bluer, similar to what it happens to LBGs. This reinforced previous claims that Ly$\alpha$ selection is strongly biased and might miss a significant percentage of the SF population at the peak of cosmic star formation.

As commented above, the $U$ band might be affected by the uncertain strength of the Ly$\alpha$ emission. The top-left inset plot in Figure \ref{figure_median_SEDs} shows the region of the SED around the Ly$\alpha$ emission. It can be seen that the $U$-band magnitudes of HAEs, LBGs, and $sBzK$ galaxies agree well with the best-fitted templates. Therefore, in these galaxies the equivalent width of the Ly$\alpha$ emission is not high enough to alter significantly the $U$-band fluxes. This indicates that HAEs, LBGs, and $sBzK$ galaxies do not have Ly$\alpha$ emission with high equivalent width on average. This is not the case for LAEs. The median $U$-band flux of LAEs is brighter than the predicted by the templates, suggesting intense Ly$\alpha$ emission with high equivalent widths as expected by their selection. These results indicate that the Ly$\alpha$ emission has high equivalent width only in a small sample of galaxies, reducing the chances of finding strong Ly$\alpha$ emission in a general population of SF galaxies at $z \sim 2$. Our selected LBGs have higher rest-frame UV luminosities than the other galaxies analysed in this work. Their lack of strong Ly$\alpha$ emission, on average, is thus in agreement with the results reported in \cite{Schaerer2011}, who found that, at a given redshift, the Ly$\alpha$ emission is more common in galaxies with fainter UV magnitudes \citep[see also][]{Stark2010MNRAS.408.1628S}.

It should be pointed out that the comparison between the different samples presented in this work is very dependent on the depth of the surveys used to select them. However, COSMOS is one of the deepest sets of multi-wavelength data available. Therefore, it makes a good judgement of the biases that affect samples selected at $z \sim 2$ through the studied selection criteria for most deep extragalactic surveys. One alternative to overcome this limitation would be using the lensing power of massive clusters of galaxies, as was done in \cite{Alavi2014ApJ...780..143A}, allowing to detect galaxies much fainter than previous surveys at $z \sim 2$. However, the latter is not the traditional way of selecting LAEs, $sBzK$, or HAEs as normally they are searched and analysed in well-known cosmological fields where a wealth of deep multi-wavelength observations are available for their analysis.

\section{Matched Ly$\alpha$ and H$\alpha$ emitters: The Ly$\alpha$ escape fraction at $z \sim 2.23$}\label{matched_HAEs_LAEs}

Of special interest is the analysis of the galaxies which exhibit both Ly$\alpha$ and H$\alpha$ emission (Ly$\alpha$-H$\alpha$ emitters, LAHAEs), not only due to the low number of such sources reported so far at $z \sim 2$, but also because they allow to constrain the escape fraction of Ly$\alpha$ photons \citep{Hayes2010Natur.464..562H,Song2014ApJ...791....3S}. As indicated in \S \ref{source_selection}, we can study LAHAEs because the Ly$\alpha$ and H$\alpha$ NB filters used in \cite{Nilsson2009} and \cite{Sobral2013MNRAS.428.1128S} select galaxies within an overlapping redshift slice over the same region of the sky. The Ly$\alpha$ observations select galaxies over the redshift range $2.206 \lesssim z_{\rm{Ly\alpha}} \lesssim 2.312$. The H$\alpha$ observations cover $2.216 \lesssim z_{\rm H \alpha} \lesssim 2.248$. The Ly$\alpha$ observations cover a wider redshift range than the H$\alpha$ observations. Therefore, if HAEs had strong Ly$\alpha$ emission, they should have been detected in the Ly$\alpha$ filter.

In the overlapping region of the COSMOS field observed by the Ly$\alpha$ and H$\alpha$ NB filters there are 158 HAEs and 146 LAEs\footnote{We should point out that we include here the whole sample of LAEs, both detected and undetected in near-IR wavelengths. The reason is that we are interested here in the fraction of HAEs with Ly$\alpha$ emission, not in the SED-derived properties of galaxies with Ly$\alpha$ emission (for which only near-IR-detected LAEs should be included -- see \S \ref{source_selection}). The seven LAHAEs have detection in near-IR wavelengths, so good SED fits can be carried out.}. However, there are only 7 galaxies in common between the two samples. Therefore, only 4.5\% of HAEs have strong enough Ly$\alpha$ emission to be selected as LAEs. This implies very low Ly$\alpha$ escape fraction at $z \sim 2$ and agrees with \cite{Hayes2010Natur.464..562H}, who found six LAHAEs in their sample, implying an average Ly$\alpha$ escape fraction of $\sim$ 5\%. These results reinforce the different nature of the galaxies selected through the H$\alpha$ and Ly$\alpha$ NB techniques and highlight the low chance of finding galaxies with Ly$\alpha$ emission, at least at $z \sim 2.23$. It should be remarked that the Hayes et al. results are based on Ly$\alpha$ observations about 10 times deeper than those used in this work. Similarly, the H$\alpha$ survey in \cite{Hayes2010Natur.464..562H} is about 2 times deeper than HiZELS. However, we cover in this work an area of the sky that is about 20 times higher than in \cite{Hayes2010Natur.464..562H}. Therefore, the results complement each other in different regimes of Ly$\alpha$ and H$\alpha$ brightness and area surveyed. We have explicitly indicated in Figure \ref{color_bzk_figure} the location of LAHAEs. Among the seven joint detections, six have blue $B - z$ colors compatible with almost flat UV continuum, whereas one of them has a red SED. Apart from the red LAHAE with SED-derived $E_s (B - V) \sim 0.5$, the remaining LAHAEs have a median dust attenuation of $E_s (B - V) \sim 0.15$. Their ages range within 1 -- 900 Myr. 

\begin{figure*}
\centering
\includegraphics[width=0.49\textwidth]{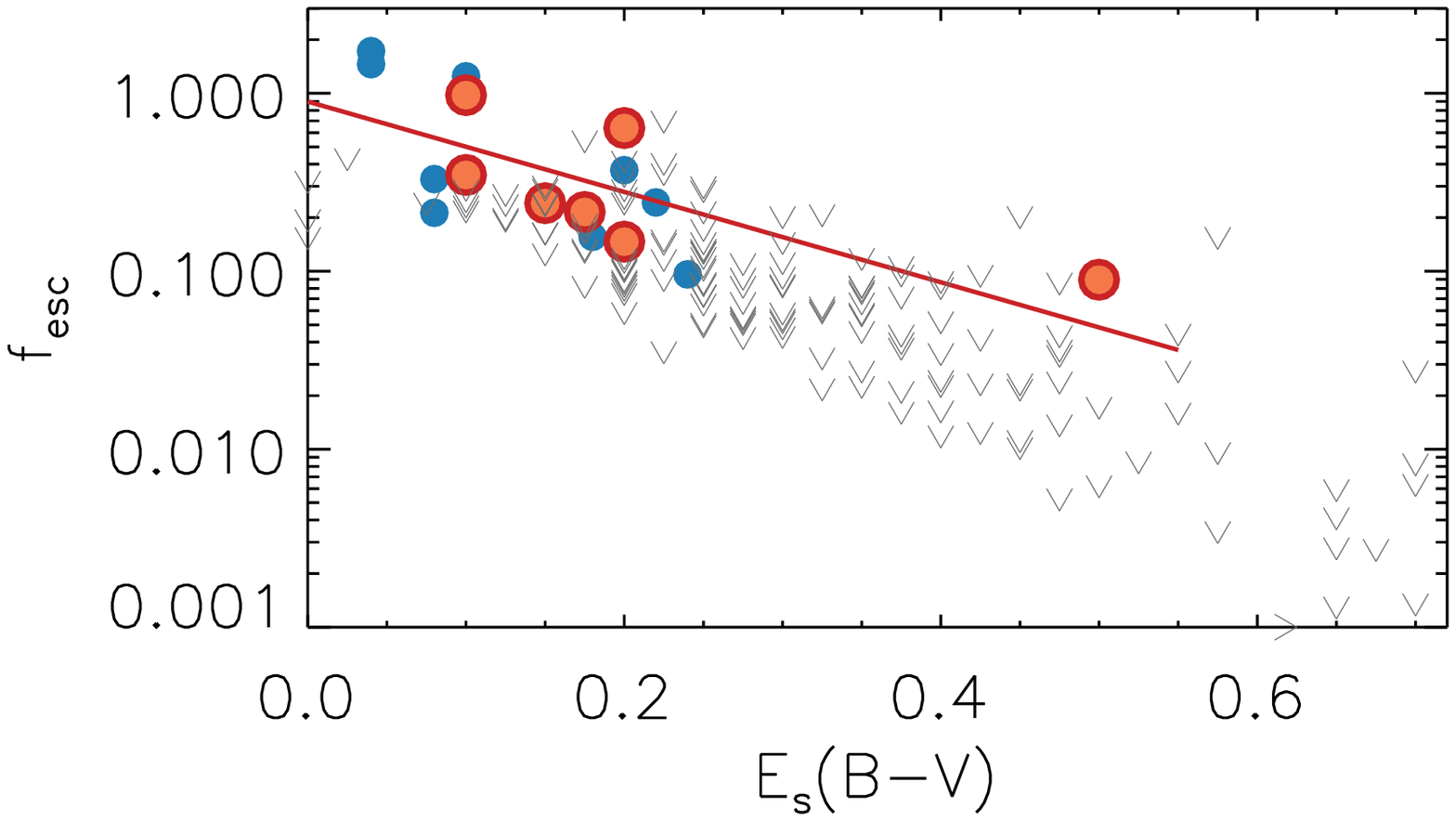}
\vspace{-7mm}
\hspace{-27mm}
\includegraphics[width=0.49\textwidth]{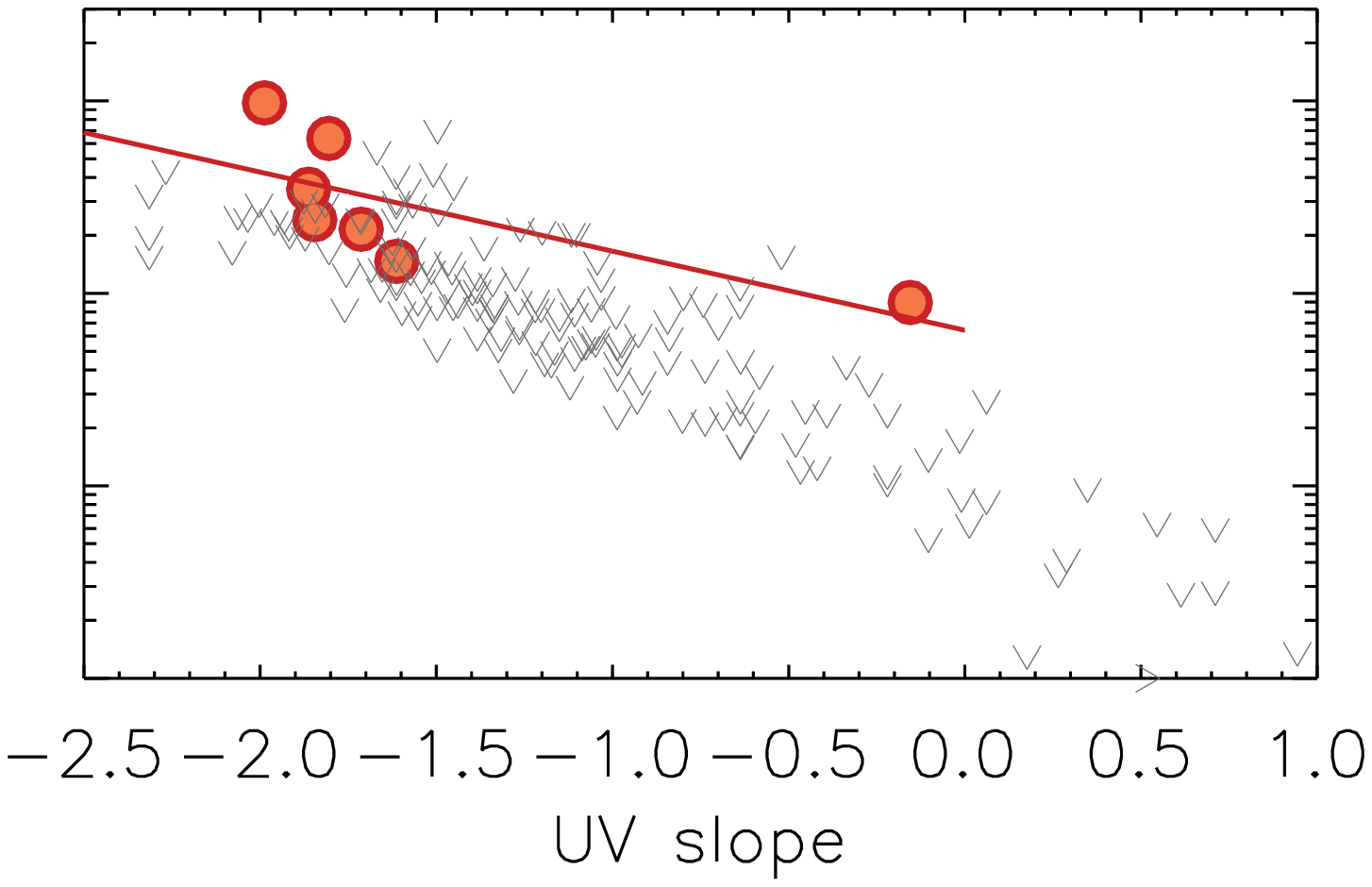}
\includegraphics[width=0.49\textwidth]{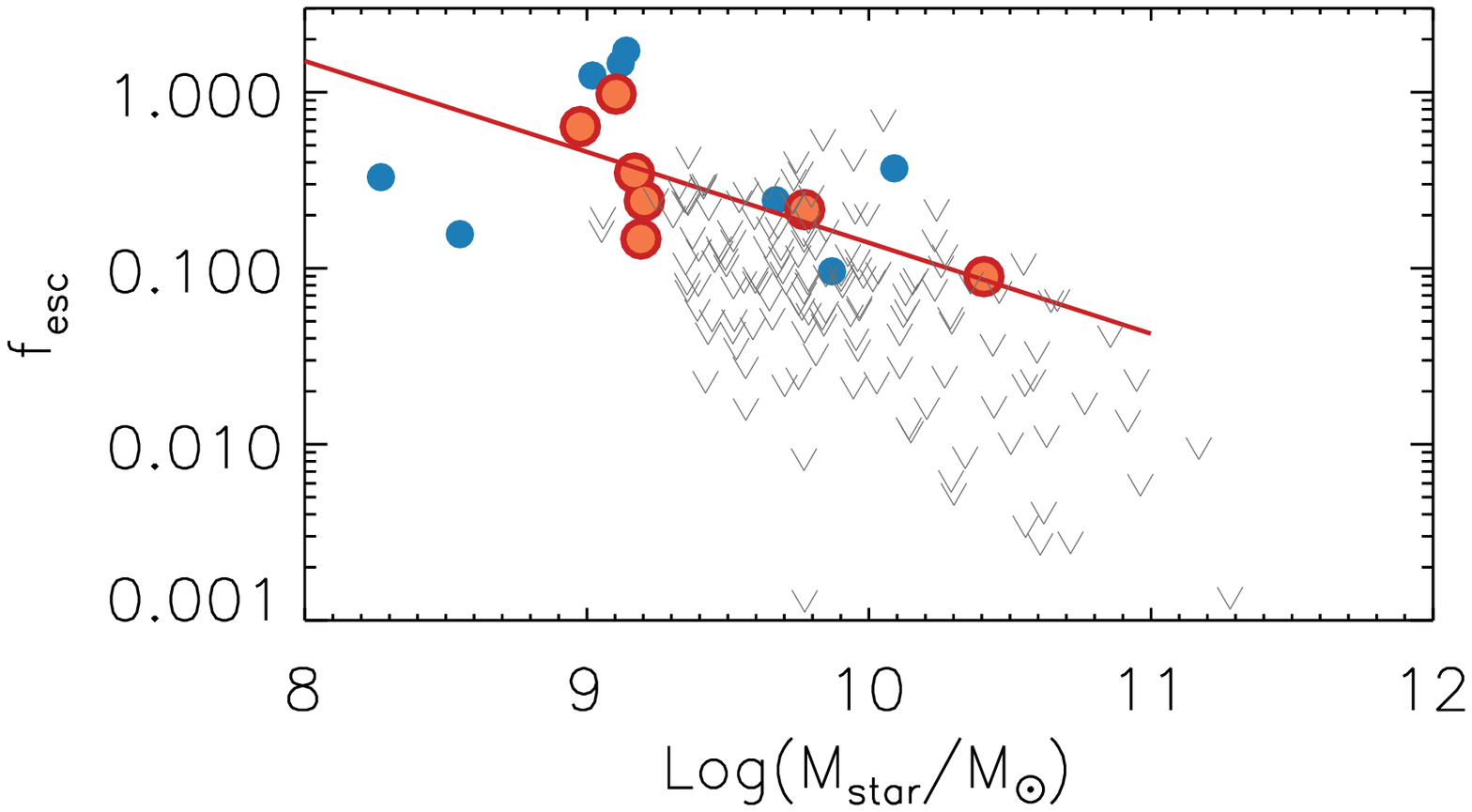}
\hspace{-27mm}
\includegraphics[width=0.49\textwidth]{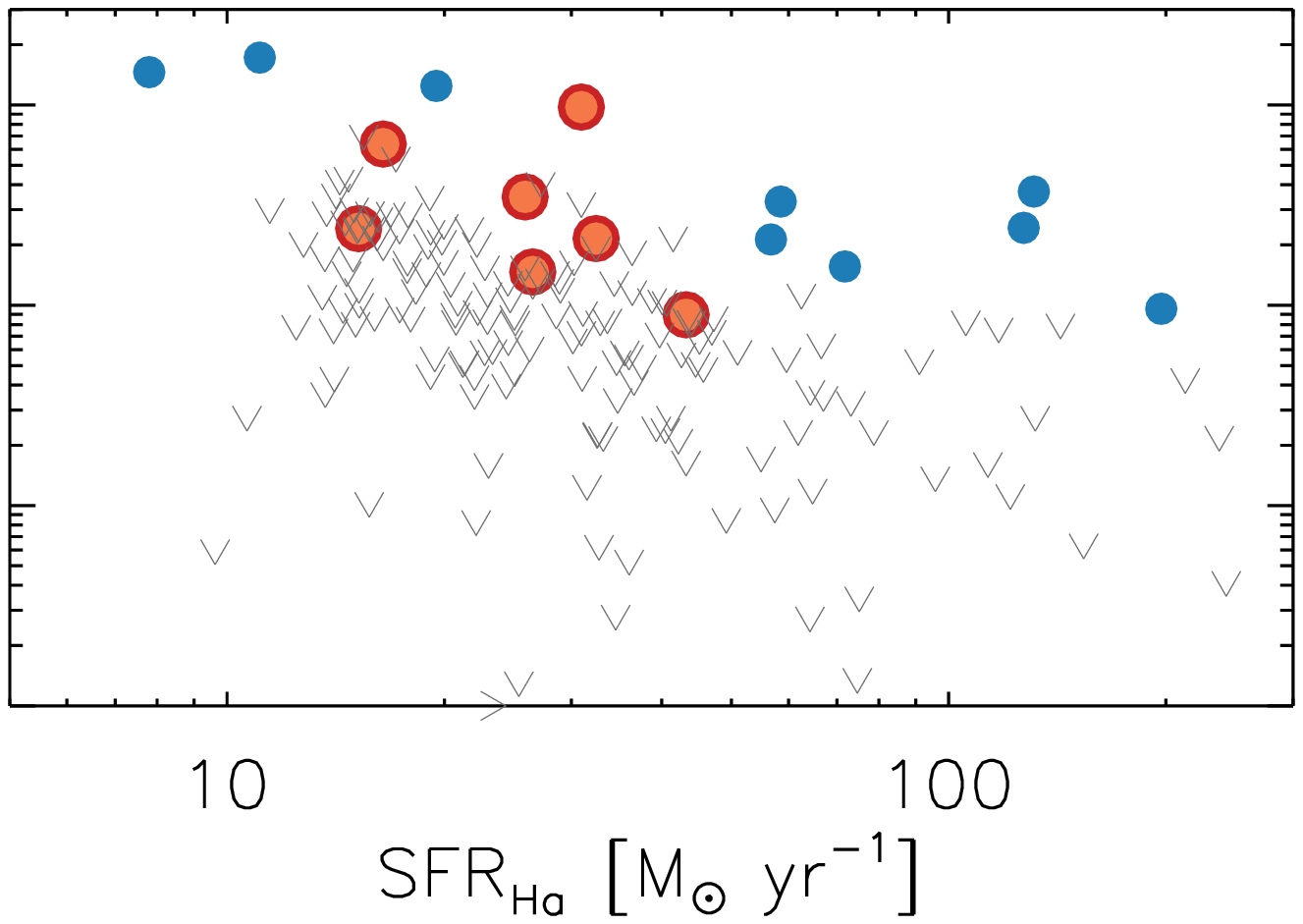}
\caption{Ly$\alpha$ escape fraction as a function of dust attenuation (\emph{upper left}), UV continuum slope (\emph{upper right}), stellar mass (\emph{bottom left}), and dust-corrected ${\rm SFR_{H\alpha}}$ (\emph{bottom right}) for our sample of LAHAEs (orange filled dots). For a comparison, we also show the nine galaxies with both Ly$\alpha$ and H$\alpha$ emission in \citet{Song2014ApJ...791....3S} with blue filled dots (UV continuum slopes are not provided for individual galaxies in \citealt{Song2014ApJ...791....3S}). The Ly$\alpha$ escape fraction has been calculated by using the SED-derived dust attenuation and assuming that $E_s(B-V) = E_g(B-V)$. This assumption is the main factor affecting the uncertainties of the results, not only in this work, but also in previous results in the literature since the relation between $E_s(B-V)$ and $E_g(B-V)$ at $z \sim 2$ has not been accurately established yet. The straight lines are linear fits to our points. We also include upper limits for the HAEs undetected in Ly$\alpha$. It can be seen that the Ly$\alpha$ escape fraction decreases with increasing dust attenuation, redder UV continuum slopes and higher stellar masses and SFRs.
              }
\label{fescape_LAEs}
\end{figure*}

We derive the Ly$\alpha$ escape fraction in our LAHAEs from the ratio between the intrinsic and observed Ly$\alpha$ luminosities and assuming case B recombination, so the intrinsic Ly$\alpha$ emission can be obtained from the intrinsic (dust-corrected) H$\alpha$ luminosity. In order to make a fair comparison with previous works, the H$\alpha$ luminosity is corrected for dust attenuation by using the SED-derived $E_s(B-V)$:

\begin{equation}
	f_{\rm esc} = \frac{L_{\rm obs} (Ly\alpha)}{8.7 \times L_{\rm obs} (H\alpha) \times 10^{0.4 \times E_s(B-V) \times k(\lambda_{\rm H\alpha})}}
\end{equation}

\noindent In the previous equation we have assumed that $E_s(B-V) = E_{g} (B-V)$, as it has been reported in several previous works to happen at high redshift \citep[see for example][]{Reddy2004ApJ...603L..13R,Erb2006ApJ...647..128E,Reddy2010}. However, it should be remarked that this has not been proven accurately, and will represent one of the major sources of uncertainties in our derived Ly$\alpha$ escape fraction. This uncertainty affects not only our results but also the results reported in most previous works at similar redshifts, since a relation between $E_{s}(B-V)$ and $E_{g}(B-V)$ must be assumed.

In Figure \ref{fescape_LAEs} we represent the Ly$\alpha$ escape fraction as a function of the SED-derived dust attenuation, UV continuum slope, stellar mass, and dust-corrected $\rm{SFR_{H\alpha}}$ for our LAHAEs at $z \sim 2.23$. For a comparison we also represent the nine LAHAEs reported in \cite{Song2014ApJ...791....3S} with both Ly$\alpha$ and H$\alpha$ emission (UV continuum slopes are not provided for individual galaxies in \citealt{Song2014ApJ...791....3S}). To be consistent, we take the observed Ly$\alpha$ and H$\alpha$ fluxes and the SED-derived properties from \cite{Song2014ApJ...791....3S} and then the Ly$\alpha$ escape fraction is calculated with the previous equation. However, we note that the selection criteria are not the same in both samples. Here we use the classical NB technique to look for LAEs and HAEs, whereas the LAEs in \cite{Song2014ApJ...791....3S} were found through blind spectroscopy and then followed-up with near-IR spectroscopic observations. Actually, the Ly$\alpha$ luminosities of the LAHAEs in \cite{Song2014ApJ...791....3S} are higher than for the LAHAEs presented in this work. Despite these differences, the Ly$\alpha$ escape fractions derived in both works agree well. 

It can be seen from Figure \ref{fescape_LAEs} that the Ly$\alpha$ escape fraction decreases with increasing dust attenuation, as previously reported at similar and lower redshifts \citep{Hayes2014ApJ...782....6H,Atek2014A&A...561A..89A,Song2014ApJ...791....3S}, and also with increasing (redder) UV continuum slope (a proxy for dust attenuation) and higher stellar mass and SFR. We also include in Figure \ref{fescape_LAEs} the upper limits on the Ly$\alpha$ escape fraction in those HAEs whose Ly$\alpha$ is not detected. These upper limits have been calculated assuming a limiting Ly$\alpha$ EW of 80 \AA\ \citep{Nilsson2009}. The upper limits indicate that the relations found for LAHAEs corresponds to the upper envelope of the actual correlations. However, upper limits also indicate that the highest Ly$\alpha$ escape fraction seen at low dust attenuation and/or UV continuum slopes are not seen at higher dust attenuation and/or UV continuum slopes. These results indicate that Ly$\alpha$ emission preferentially escapes from blue, low-mass galaxies with low dust attenuation (with the exception of a low percentage of massive, red and dusty LAEs as obtained in \S \ref{comparing_things}). This is in agreement with the SED-derived properties and colors found for LAEs in previous sections. It should be noted that the number of LAHAEs at $z \sim 2$ reported so far is low and, consequently, matched Ly$\alpha$-H$\alpha$ surveys over larger areas of the sky are needed to increase the number of LAHAEs and have more robust results with higher statistical significance.

\begin{figure*}
\centering
\includegraphics[width=0.49\textwidth]{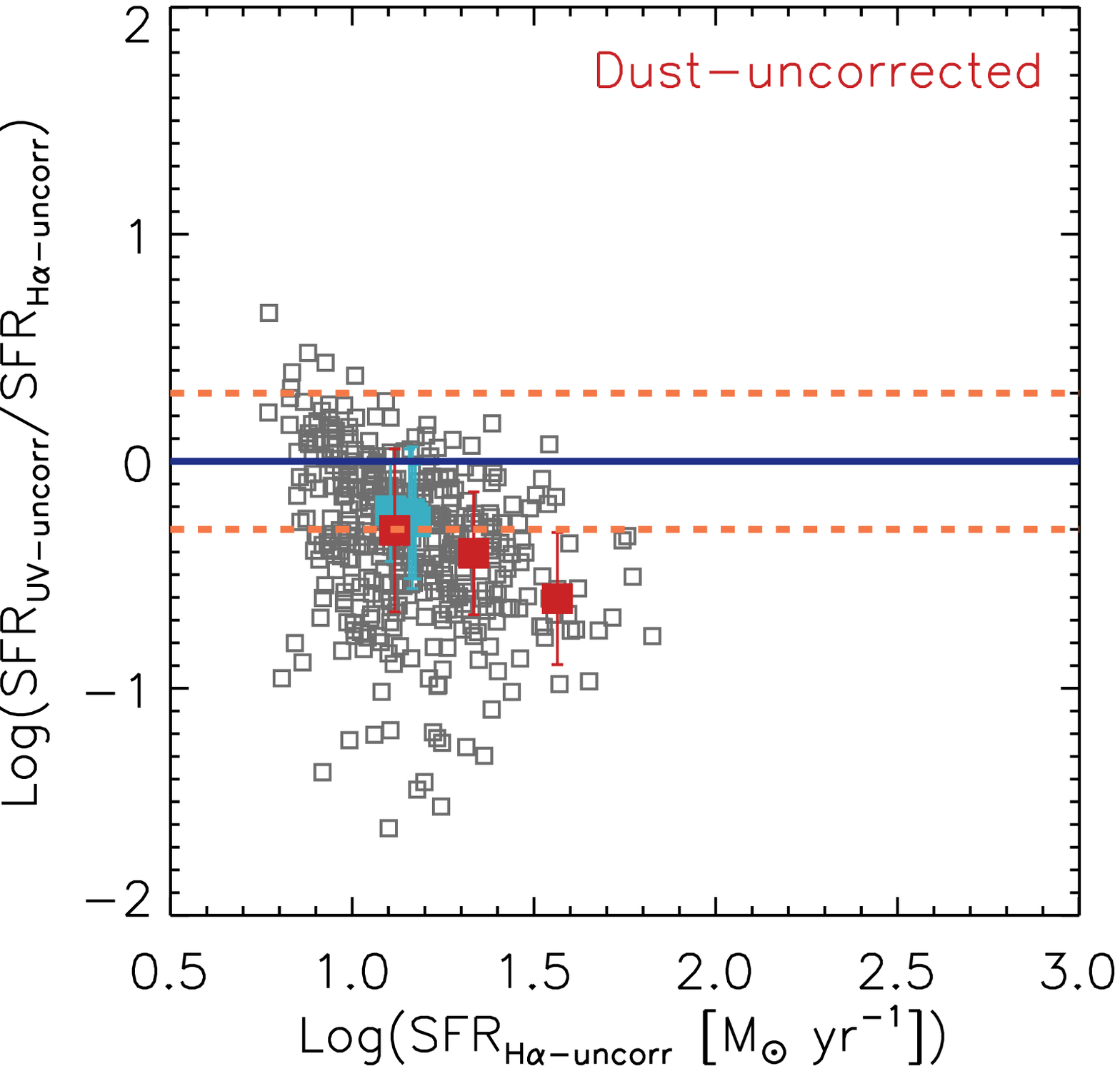}
\includegraphics[width=0.49\textwidth]{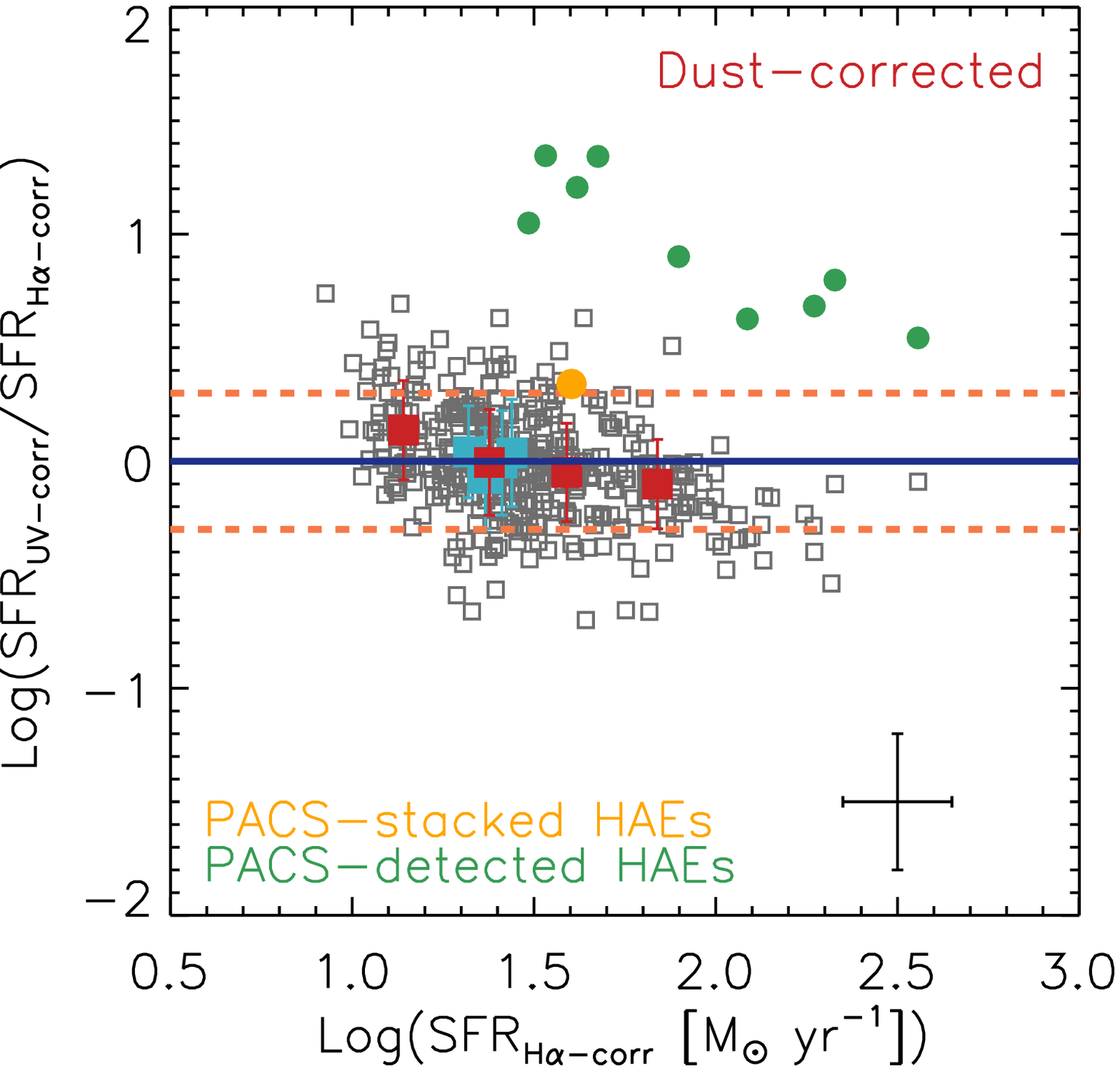}
\caption{Comparison between the SFR derived from H$\alpha$ and rest-frame UV luminosities for our studied HAEs at $z \sim 2$. \emph{Left panel}: results without dust correction. \emph{Right panel}: results when dust correction is included. The H$\alpha$ luminosities are corrected by dust attenuation by using the local relation between dust attenuation and stellar mass \citep{Garn2010MNRAS.409..421G}, suggested to be valid at least up to $z \sim 1.5$ \citep{Sobral2012MNRAS.420.1926S}. The rest-frame UV luminosities are corrected by using the dust attenuation derived with the \citet{Heinis2013} relation. In both panels, the filled red squares represent the median SFR$_{\rm UV}$ in different bins of SFR$_{\rm H \alpha}$ and filled cyan squares represent the median SFR$_{\rm UV}$ in different bins of stellar mass. The solid lines are the one-to-one relations and the dashed lines are deviations of $\pm 0.3$ dex. PACS-detected and stacked HAEs are represented in the right panel by filled green and orange dots, respectively. The error bars on the right panel shows a lower limit on the uncertainties of the results. For the dust correction applied to H$\alpha$ luminosity, this uncertainty comes from the scatter in the best-fitted relation found in \citet{Sobral2012MNRAS.420.1926S}. It should be pointed out that the uncertainties affecting to the dust corrections in H$\alpha$ are much lower than those affecting to the rest-frame UV continuum, since the latter is much more affected by dust.
              }
\label{sfr_comparison_UV}
\end{figure*}

\section{H$\alpha$ SFR vs UV and UV+IR: a consistent view}\label{sfr_ha_good}

The H$\alpha$ emission is an excellent tracer of instantaneous star formation and calibrations between the SFR and the H$\alpha$ luminosity have been proposed in the literature \citep[see for example][]{Kennicutt1998}. The H$\alpha$ emission is affected by dust attenuation (although to a much lesser extent than, for example, the UV). Therefore, in order to derive the total SFR from H$\alpha$, accurate dust correction factors are needed. In the local universe, where a significant percentage of galaxies can be detected in the FIR or their H$\alpha$ and H$\beta$ emissions can be measured, the dust correction factors can be determined with acceptable accuracy. However, this is much more challenging for galaxies in the high-redshift Universe, complicating the determination of the total SFR. 

We examine in this section the robustness of the H$\alpha$ emission as a tracer of SFR at $z \sim 2.23$ for our HAEs. We first compare the results obtained from H$\alpha$ and UV estimates of the SFR when dust correction is not taken into consideration. This is shown on the left panel of Figure \ref{sfr_comparison_UV}. As it can be seen, the two tracers of star formation give significantly different results, with the H$\alpha$-derived SFR being higher than the obtained from the rest-frame UV luminosity. The median values found are ${\rm SFR_{H\alpha}} = 13.7 \pm 9.0 \, {\rm M}_{\odot} {\rm yr}^{-1}$ and ${\rm SFR_{\rm UV}} = 7.4 \pm 6.6 \, {\rm M}_{\odot} {\rm yr}^{-1}$, where the uncertainties represent the interquartile ranges.

In the right panel of Figure \ref{sfr_comparison_UV} we show the comparison between H$\alpha$-derived and UV-derived SFR when dust correction is included. We correct the rest-frame UV luminosities by using the relation between the dust attenuation and UV continuum slope derived in \cite{Heinis2013}, as in \S \ref{comparing_things}. The dust correction of the H$\alpha$ luminosity has been derived by using the relation between stellar mass and dust attenuation of \cite{Garn2010MNRAS.409..421G}, that has been suggested to be valid for HAEs at least up to $z \sim 1.5$ \citep{Sobral2012MNRAS.420.1926S,Ibar2013MNRAS.434.3218I,Dominguez2013ApJ...763..145D,Price2014ApJ...788...86P}. In the right panel of Figure \ref{sfr_comparison_UV}, it can be seen that there is a very good agreement between the UV-derived and the H$\alpha$ derived SFRs, despite these being calculated with completely different and independent estimators at different wavelengths. The difference between the two estimators is typically lower than 0.3 dex for individual galaxies. Furthermore, both estimations agree quite well when galaxies are averaged over different bins of H$\alpha$-derived SFR or stellar mass. This result shows the robustness of the H$\alpha$ emission as tracer of SFR at $z \sim 2$, which adds to the unbiased and well-understood selection function of the H$\alpha$ NB method to find SF galaxies at different redshifts \citep{Sobral2012MNRAS.420.1926S}.

Previous works have also analysed the accuracy of H$\alpha$ emission to recover SFR at high redshift, for example \cite{Erb2006ApJ...647..128E} and \cite{Reddy2010}. They correct from dust attenuation derived from SED fitting, $E_s(B-V)$, and assuming $E_s(B-V) = E_g(B-V)$, where $E_g(B-V)$ is the reddening for nebular emission. They actually obtained that the traditionally employed relation $E_s(B-V) = 0.4 \times E_g(B-V)$ \citep{Calzetti2000} produces H$\alpha$-SFR that over-predict those derived from the X-ray and dust-corrected UV emissions. Recently, \cite{Steidel2014ApJ...795..165S} apply a relation between $A_{\rm H\alpha}$ and $E_s(B-V)$ that depends on the value of $E_s(B-V)$. The main advantage of our dust correction method for H$\alpha$ emission is that we do not require any assumptions about the relationship between $E_s(B-V)$ and $E_g(B-V)$, which is still until debate and have been suggested to be redshift-dependent \citep{Kashino2013ApJ...777L...8K}. Instead, we use a relation between dust attenuation and stellar mass that has been reported to be valid at least up to $z \sim 1.5$ for HAEs \citep{Sobral2012MNRAS.420.1926S}.

We also compare in the right panel of Figure \ref{sfr_comparison_UV} the H$\alpha$-derived total SFR with the total SFR obtained with direct \emph{Herschel} detections for the 9 PACS/SPIRE-detected HAEs. As it can be seen, the H$\alpha$ emission, after the dust correction has been included, is not able to recover the more accurate SFR derived with PACS detections. This is because, at $z \sim 2$, \emph{Herschel} only detects the most extreme sources for which our average dust correction factor applied to the H$\alpha$ luminosity are not high enough (due to very high  internal obscuration) to reproduce the more accurate \emph{Herschel}-derived SFR. This also happens to the dust correction factors derived from the UV continuum slopes in high-redshift \emph{Herschel}-selected galaxies \citep[see for example][]{Oteo2013A&A...554L...3O, Rodighiero2014MNRAS.443...19R}. Stacking as a function of the H$\alpha$-derived total SFR we only recover one $> 3 \sigma$ stacked detection, also represented in the right panel of Figure \ref{sfr_comparison_UV}. The total SFR derived from the PACS stacked flux is higher (in about $\sim$ 0.3 dex) with respect to the H$\alpha$-determination. This might indicate that the dust-correction factor used to correct the H$\alpha$ emission is more uncertain for the most massive HAEs with the highest SFRs, although deeper FIR data would be needed to confirm this trend to more normal SF galaxies with lower SFR.

\section{SFR versus stellar mass relation and its uncertainties at $z \sim 2$}\label{sfr_mass_haes_MS}

\subsection{The main sequence for HAEs at $z \sim 2$}\label{reimdoVAVAV}

\begin{figure}
\centering
\includegraphics[width=0.49\textwidth]{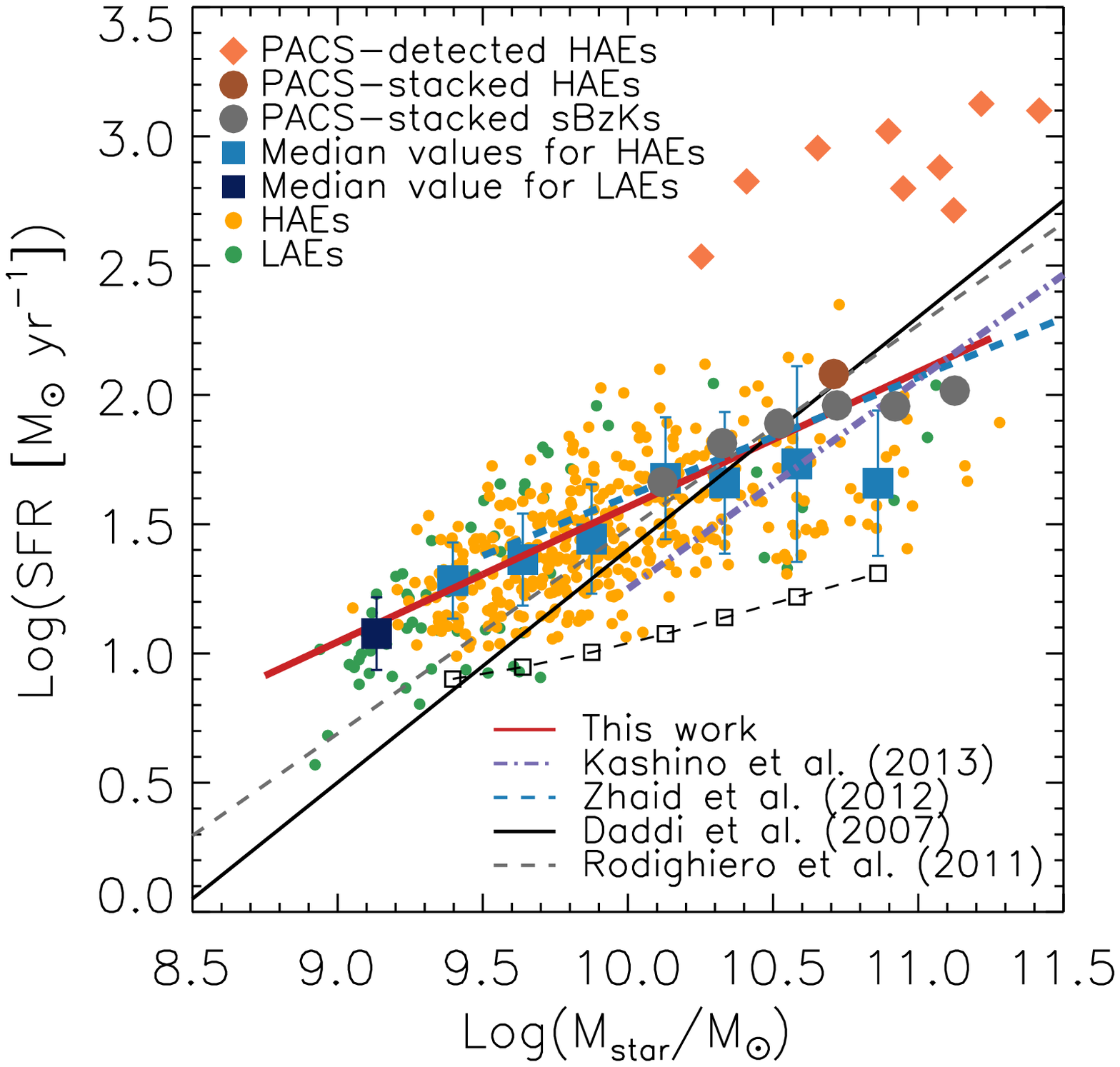}
\caption{Relation between the SFR and stellar mass for our HAEs at $z \sim 2.23$ (orange dots). Light blue squares represent the median SFR in different bins of stellar mass. The SFR of HAEs is derived with the \citet{Kennicutt1998} calibration and the H$\alpha$ luminosity corrected from dust attenuation with the local dust-mass relation. The location of \emph{Herschel}-detected and stacked HAEs is indicated with the orange diamonds and the brown filled dot, respectively, and their total SFR are ${\rm SFR_{total}} = {\rm SFR_{UV}} + {\rm SFR_{IR}}$. For comparison, we also show a sample of stacked $sBzK$ galaxies with filled grey dots. LAEs are represented with green dots and their SFR are derived from the dust-corrected rest-frame UV luminosity with the \citet{Heinis2013} relation. Previously reported MS at $z \sim 2$ are over-plotted, as indicated in the bottom-right legend. For all the galaxies, the stellar masses have been calculated from SED fitting with BC03 templates including emission lines. The open black squares and dashed curve represent the limiting dust-corrected SFR as a function of mass for our H$\alpha$ survey (see text for more details). It can be seen from this figure that the HAEs follow an MS with a lower slope than the MS at $z \sim 2$ reported in previous works. Furthermore, our results indicate that there is a flattening of the MS at the highest stellar masses, as previously reported in previous works from different analysis based on different samples of SF galaxies.
              }
\label{sfr_mass_Halpha}
\end{figure}

Most previous works agree that there is a relation between SFR and stellar mass, commonly referred to as the main sequence (MS), where normal SF galaxies are located (see \citealt{Speagle2014ApJS..214...15S} for a recent compilation). The MS has been reported to exist from the local universe up to high-redshift and to be relatively independent of environment \citep{Koyama2013MNRAS.434..423K}. However, less is known about the values of its zero point and slope at a given redshift. One of the main reasons is the lack of FIR detections for a representative population of SF galaxies at each redshift, that prevents accurate determinations of the total SFR. Even with the deepest FIR surveys carried out so far \citep[see for example][]{Elbaz2011}, only a very small fraction of the galaxies are detected in the FIR, mostly at the highest redshifts \citep{Oteo2013A&A...554L...3O,Oteo2014MNRAS.439.1337O}. Furthermore, it is claimed that most high-redshift FIR-detected galaxies are not normal SF galaxies, but likely have a starburst nature and are preferentially located above the MS \citep{Rodighiero2011ApJ...739L..40R,Lee2013ApJ...778..131L,Oteo2013A&A...554L...3O,Oteo2014MNRAS.439.1337O}. Therefore, even with \emph{Herschel}-selected galaxies it is not possible to determine the slope and zero point of the MS at high redshift. 

Now that we have evidence from \S \ref{sfr_ha_good} that the H$\alpha$ emission is a good tracer of star formation at $z \sim 2.23$, that H$\alpha$ provides a clean, well-understood, SFR-selected sample (not biased to either just blue or just red galaxies), that HAEs are an excellent representation of the whole population of SF galaxies at $z \sim 2$ (\S \ref{comparing_things}), and that their stellar masses cover a wider range than many previous studies at $z \sim 2$, we can attempt to study the relation between SFR and stellar mass and its uncertainties by using our sample of HAEs. This is shown as the orange points in Figure \ref{sfr_mass_Halpha}, with the best-fit MS indicated by the red line.

The main caveat of the analysis of the MS with our sample of HAEs is that they are selected down to a fixed ${\rm SFR_{H\alpha-uncorr}}$ (${\rm \sim 3 \, M_\odot \, \rm yr^{-1}}$ and EW cut, \citealt{Sobral2013MNRAS.428.1128S}). Furthermore, the ${\rm SFR_{H\alpha-corr}}$ of HAEs is obtained by using a dust correction factor which depends on the relation between dust attenuation and stellar mass. All of this means that, for a given stellar mass, only HAEs with ${\rm SFR_{H\alpha-corr}}$ above a certain limit (that is dependent on stellar mass) are included in the sample. This might produce a flattening of the MS with respect to previous works based on other selection functions. In order to explore this in more detail, we have obtained the limiting ${\rm SFR_{H\alpha-corr}}$ in different bins of stellar mass. This is represented in Figure \ref{sfr_mass_Halpha} with the open black squares and the dashed curve. As expected, the limiting curve tracks along the locus of the bottom of the points corresponding to HAEs. As an example, our H$\alpha$ selection would not be able to detect galaxies on the MS relation of \citet[solid black line on Figure \ref{sfr_mass_Halpha}]{Daddi2007} that have stellar masses $\log{\left( M_*/M_\odot \right) \sim 9.5}$ (but note that dust correction factors used to recover total SFR are different in \cite{Daddi2007} and this work). If our H$\alpha$ observations would have selected galaxies down to a lower ${\rm SFR_{H\alpha-uncorr}}$, we could have been able to select galaxies with lower ${\rm SFR_{H\alpha-corr}}$ at that stellar mass and, consequently, we might have obtained a steeper MS. This reflects the effect of our H$\alpha$ selection on the slope of the MS at $z \sim 2$ and should be taken into account in any comparison with the results presented in the literature for galaxies selected in a different way. However, as it will be shown in the next Section, when we use the same methods that in other works to determine the ${\rm SFR_{H\alpha-corr}}$ and stellar mass, the MSs are in very good agreement.

It is clear from Figure \ref{sfr_mass_Halpha} that an MS for HAEs exists at $\log{\left( M_*/M_\odot \right)} > 9.25$: more massive HAEs have higher SFRs on average. However, the scatter is significant and it increases with stellar mass. It can be also seen that the MS is flatter for more massive galaxies. Actually the relation becomes almost flat at $\log{\left( M_* / M_\odot \right)} \geq 10.2$, where the width of the MS is also higher. Since HAEs are selected by star formation, they are truly SF galaxies: all HAEs with $\log{\left( M_* / M_\odot \right)} \geq 10.0$ would also have been selected as SF galaxies according to the $sBzK$ criterion. Therefore, the flattening is not caused by the presence of massive quiescent galaxies without star formation. This flattening of the MS towards massive galaxies can be also seen in \cite{Heinis2014MNRAS.437.1268H} at $z \sim 1.5$ in their work about stacking analysis in \emph{Herschel} bands, although the flattening found in this work is much more significant (see also \citealt{Oteo2014A&A...572L...4O}, \citealt{Whitaker2014ApJ...795..104W}, or \citealt{Schreiber2014arXiv1409.5433S}). By fitting a linear relation for galaxies with $\log{\left( M_* / M_\odot \right)} \leq 10.25$ in the form $\log{\left( \rm SFR_{H\alpha -corr}\right) } = a + b \times \log{\left( M_*/M_\odot\right)}$ we find $a = -3.65 \pm 0.15$ and $b= 0.52 \pm 0.02$. This MS is in very good agreement with the results shown in \cite{Zahid2012ApJ...757...54Z} in a sample of spectroscopically confirmed H$\alpha$ emitters over a similar stellar mass range than our sample. This is the most comparable sample to ours that can be found in the literature in terms of both H$\alpha$ emission and stellar mass range. \cite{Kashino2013ApJ...777L...8K} obtained a higher MS slope with a spectroscopically confirmed sample of galaxies with H$\alpha$ emission, but their galaxies are at a slightly lower median redshift than HAEs in this work and their sample is restricted to more massive galaxies, mainly with $\log{\left( M_*/M_\odot \right)} > 10$. Our MS has a lower slope than the classical MS found in \cite{Daddi2007} (D07), probably produced by our selection effect, although different dust-correction factors and recipes for stellar mass determination might play a significant role. This will be discussed in more detail in \S \ref{uncerSFmas}.

PACS-detected HAEs are all well above the HAE MS, as happens for PACS-detected LBGs and $sBzK$ galaxies \citep{Oteo2014MNRAS.439.1337O,Rodighiero2014MNRAS.443...19R}. This indicates that they have an SB nature and do not belong to the population of normal SF galaxies at $z \sim 2$. Actually, our PACS-detected $sBzK$ galaxies have even higher SFRs than those in \cite{Oteo2014MNRAS.439.1337O} due to the shallower PACS data in COSMOS \citep{Lutz2011}. When stacking as a function of stellar mass, we only recover one detection for HAEs, also represented in Figure \ref{sfr_mass_Halpha}. The stacked point is in agreement with the D07 MS and it is located slightly above the MS for HAEs found in this work. This is compatible with the previous result that the SFR derived from stacking is slightly higher than the derived from H$\alpha$ in the most massive HAEs. For comparison, we also show the stacked points corresponding to the sample of $sBzK$ galaxies. Only the most massive $sBzK$ galaxies are detected through stacking. The stacked points for $sBzK$ galaxies agree very well with the extrapolation of the MS for $\log{\left( M_* / M_\odot \right)} \leq 10.25$ HAEs towards higher stellar masses. However, the most massive stacked $sBzK$ galaxies have higher SFR than the most massive HAEs. This difference might be due to the fact that the dust correction factor employed to recover the total SFR might not be accurate for the most massive galaxies. However, it could also be as a consequence of a different nature between HAEs and $sBzK$ galaxies for the most massive objects. The latter would be supported by the shape of their UV-to-NIR SEDs (see Figure \ref{figure_median_SEDs}), since massive HAEs are bluer than massive $sBzK$ galaxies.

As already obtained in \S \ref{comparing_things}, galaxies selected through the Ly$\alpha$ technique tend to be less massive than HAEs. Furthermore, the Ly$\alpha$ selection allows to probe down to lower dust-corrected SFRs. This can be used to extend the MS obtained for HAEs down to $\log{\left( M_* / M_\odot \right)} \sim 9$. This is represented by the blue filled squared in Figure \ref{sfr_mass_Halpha}. As in \S \ref{comparing_things}, the dust-corrected SFR for LAEs has been obtained by assuming the IRX-$\beta$ relation of \cite{Heinis2013}. Interestingly, and despite the effect of our H$\alpha$ selection on the MS slope, the point for LAEs agrees very well with the extrapolation of the HAE-MS to lower stellar masses. 


\begin{figure}
\centering
\includegraphics[width=0.49\textwidth]{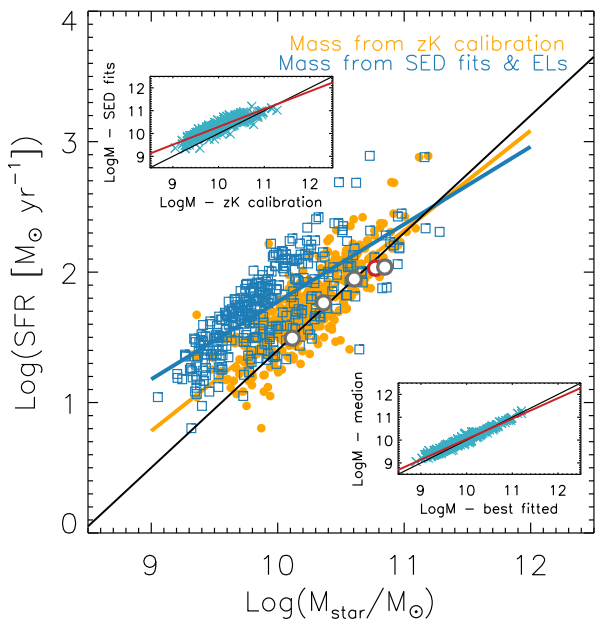}
\caption{Relation between dust-corrected total SFR and stellar mass for our HAEs when different recipes to derive stellar mass are used: SED fits with BC03 templates and inclusion of emission lines (blue open squares) and \citet{Daddi2004} calibration between stellar mass and rest-frame optical color (orange filled dots). In all cases we derive dust-corrected total SFR from the rest-frame UV luminosity corrected with the relation of \citet{Meurer1999} IRX-$\beta$. For comparison, we also show the points associated to stacked $sBzK$ galaxies (grey open dots) and HAEs (red open dot) in PACS-160 $\mu$m whose stellar masses are obtained from the \citet{Daddi2004} calibration. Inset panels show the relation between stellar masses when different approaches are considered for their calculation: SED-derived mass against masses obtained with the \citet{Daddi2004} calibration (top-left panel) and SED-derived median versus best-fitted masses (bottom-right panel). In both inset panels, black line is the one-to-one relation, and the red line is the linear fit to the points. This figure indicates that the definition of the MS (slope and normalization) depends on the way stellar mass is calculated.
              }
\label{discussion_MS_masa}
\end{figure}

\subsection{Uncertainties in the SFR-mass relation}\label{uncerSFmas}

In order to study the uncertainties involving the determination of the slope and zero-point of the HAE MS at $z \sim 2$ and try to explain in more detail the differences found with previous works, we compare now the SFR-mass relation when considering different recipes used in the literature to obtain stellar mass and total SFR. We focus first on the dependance of the HAE MS upon the recipe used for determining stellar mass. This is shown in Figure \ref{discussion_MS_masa}, where all the SFRs have been derived by correcting the rest-frame UV luminosity with the IRX-$\beta$ relation of \cite{Meurer1999} for the sake of homogeneity with previous works that will be referred to in this Section. Therefore, the main difference would be only the determination of stellar mass. We include in Figure \ref{discussion_MS_masa} the MS when the stellar mass of HAEs is obtained through our method of SED fits with BC03 templates associated to exponentially declining SFHs and the inclusion of emission lines (blue open squares, see \S \ref{SED_fitting_procedure}) and also (orange dots) that obtained using the calibration between $BzK$ photometry and stellar mass reported in \cite{Daddi2004} (see also \citealt{Rodighiero2014MNRAS.443...19R}). It can be clearly seen that the slope of the HAE MS depends upon the method used for deriving stellar mass, with the slope being lower when stellar masses are derived with SED fits with inclusion of emission lines. The top-left inset plot of Figure \ref{discussion_MS_masa} compares the stellar masses derived with BC03 templates including the effect of emission lines with the derived with the \cite{Daddi2004} calibration. It can be seen that they are correlated but do not follow the one-to-one relation, explaining the difference in the MS. 

When using the same dust correction factors and recipe for stellar mass determination than in D07, we find a MS slope closer to their result (cf. orange and black lines in Figure \ref{discussion_MS_masa}). The still slightly lower MS slope might be explained by the H$\alpha$ selection effect (see \S \ref{reimdoVAVAV}). We also plot in Figure \ref{discussion_MS_masa} the stacked points for HAEs and $sBzK$ galaxies when stellar masses are obtained as in \cite{Daddi2004}. It can be seen that they all follow quite well the \cite{Daddi2007} MS, indicating that if stellar masses are calculated in the same manner then the \cite{Daddi2007} MS is recovered from different SFR estimators \citep[see also][]{Rodighiero2014MNRAS.443...19R}. We have also compared the best-fitted stellar masses with the median values as derived by \emph{LePhare} (bottom-right panel in Figure \ref{discussion_MS_masa}). However, no significant differences are seen between both results that might significantly alter the slope of the HAE MS at $z \sim 2$.

\begin{figure}
\centering
\includegraphics[width=0.49\textwidth]{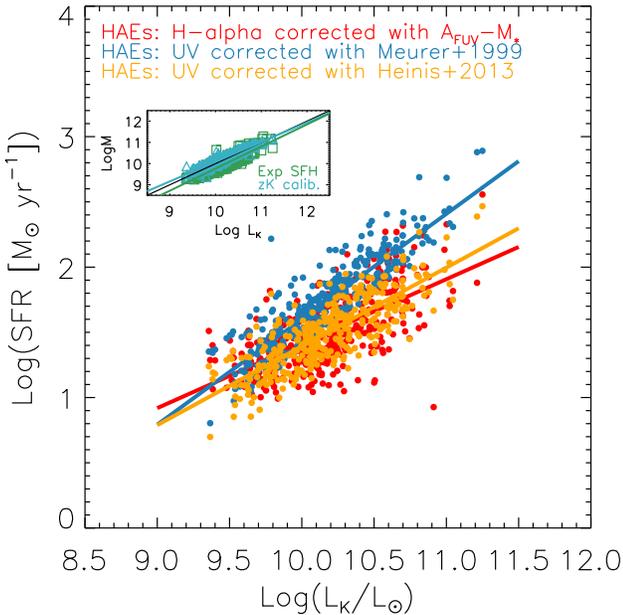}
\caption{Relation between total SFR and the rest-frame $K$ band luminosity for our sample of HAEs at $z \sim 2.23$. Different estimators of the total SFR have been used: dust-corrected H$\alpha$ luminosity with the local dust-mass relation (red dots) and dust-corrected UV rest-frame luminosity with the \citet{Heinis2013} (blue) and \citet{Meurer1999} IRX-$\beta$ (orange) relations. The inset panel shows the relation between the stellar mass and the rest-frame $K$ band luminosity for two different recipes for stellar mass: SED fits with BC03 templates (green) and calibration with $z$ and $K$ magnitudes \citep{Daddi2004} (blue). The black line would correspond to mass-to-light ratio equal to unity. Green and blue straight lines are fitted to the green and blue points, respectively. This figure indicates that dust-correction factors have a strong influence on the definition of the MS at $z \sim 2$, both in the slope and zero point.
              }
\label{discussion_MS_dust}
\end{figure}

Figure \ref{discussion_MS_dust} shows the impact of different dust correction factors in the definition of the HAE MS at $z \sim 2$. With the aim of avoiding any uncertainty coming from the determination of stellar mass we represent the SFR-mass relation by using the rest-frame $K$ band luminosity. This luminosity is a proxy of stellar mass \citep{Drory2004ApJ...608..742D} and is a direct observable quantity that does not need any assumption or calibration in its determination. Three different methods for deriving the total SFR have been employed: 1) dust-corrected H$\alpha$ luminosity with the local dust-mass relation (red points); 2) dust corrected rest-frame UV luminosity with the IRX-$\beta$ relation of \cite{Heinis2013} (orange filled dots); 3) dust corrected rest-frame UV luminosity with the IRX-$\beta$ relation of \cite{Meurer1999} (blue filled dots). Note that the only difference in Figure \ref{discussion_MS_dust} are the methods adopted for deriving the total SFR. It can be seen that the MS is strongly affected by the different dust corrections, with the slope being steeper with the dust correction based on the \cite{Meurer1999} relation. The MS derived from the H$\alpha$ emission assuming the local dust-mass relation and from the rest-frame UV with the dust correction based on the relation of \cite{Heinis2013} are similar due to the agreement between both SFR indicators (see Figure \ref{sfr_comparison_UV}). For a reference, we show in the inset panel of Figure \ref{discussion_MS_dust} the relation between stellar mass and rest-frame $K$ band luminosity for the two different assumptions for stellar mass determination used in the discussion above. 

Summarizing, our results show that, despite there is evidence that the MS exists, its slope and zero points at each redshift are challenging to determine. Their values depend on the recipes employed to obtain stellar mass and total SFR. For given assumptions of these, consistent values can be determined provided that representative samples of sources such as NB-selected HAEs or $sBzK$ galaxies are used for their characterization.

\section{Conclusions}\label{conclu}

In this work, we have carried out a multi-wavelength analysis (from the rest-frame UV to the FIR) of the spectral energy distribution (SED) of narrow-band (NB) selected, star-forming (SF) H$\alpha$ emitters (HAEs) with the aim of analysing their physical properties and their importance for galaxy formation and evolution. We have also compared their physical properties with those derived for other classical populations of SF galaxies at their same redshift: NB-selected Ly$\alpha$ emitters (LAEs), Lyman-break galaxies (LBGs), and $sBzK$ galaxies. The main conclusions of our work are the following:

\begin{enumerate}

	\item The HAE selection recovers the full diversity of SF galaxies at $z \sim 2$. Coupled with the simple and well-understood HAE selection (selecting SF galaxies down to a given dust-uncorrected SFR) which can be self-consistently applied at multiple redshift slices from the local Universe up to $z \sim 2.5$, HAEs represent an excellent sample to study galaxy evolution.
	
	\item At the depth of the COSMOS data, only about 30\% of $sBzK$ galaxies can be selected with the drop-out technique, whereas 95\% of LBGs can be selected with the $sBzK$ criterion. Only 4.5\% of HAEs can be selected as LAEs. These results indicate that the Lyman-break and Ly$\alpha$ selections miss a relevant percentage of the SF population at the peak of cosmic star formation. LBGs and LAEs are biased towards blue, less massive galaxies. Although the precise numerical results will depend on the depth of the observations considered in the analysis, these results are important to interpret results at higher redshifts where only Ly$\alpha$ or Lyman-break selections can be applied. On the other hand, most LBGs and LAEs can be selected through the $sBzK$ criterion. Only the bluest LAEs and LBGs would be missed, although this sample represents a very low percentage of the SF population at $z \sim 2$ and their exclusion would not affect significantly conclusions for galaxy evolution.
	
	\item There is a significant percentage of LAEs that are not detected in optical and near-IR broad band filters even with the deep photometry used in this work. The non detection indicates that these galaxies have a faint continuum but strong emission lines. Due to their non detection, SED fits cannot be carried out for these galaxies and, consequently, their properties and, most importantly, their contribution to galaxy evolution studies, are unknown. 
	
	\item Although the Ly$\alpha$ criterion preferentially selects SF galaxies with low dust attenuation and low stellar mass (likely due to the resonant nature of the Ly$\alpha$ emission), there is also a small percentage of red and massive LAEs in our sample, even redder than any LBG in our LBG sample. This indicates that Ly$\alpha$ is also able to escape from dusty and massive galaxies, in agreement with previous work reporting Ly$\alpha$ emission in sub-mm and FIR selected samples. Since the number of red LAEs is low compared with the total population, surveys over wider areas are needed to study in detail the properties of these galaxies and shed light on how Ly$\alpha$ can escape from these systems.
	
	\item The median SEDs of the galaxies studied in this work reveal that the Ly$\alpha$ emission is strong enough to affect the broad-band $U$ photometry only in LAEs. This result indicates that Ly$\alpha$ emission is not strong in LBGs, HAEs, or $sBzK$ galaxies on average, highlighting the low probability of finding Ly$\alpha$ emission in a general sample of selected SF galaxies.
	
	\item Only 4.5\% of HAEs show detectable Ly$\alpha$ emission implying low Ly$\alpha$ escape fraction at $z \sim 2$ in agreement with previous results. Additionally, we find that the Ly$\alpha$ escape fraction ($f_{\rm esc}$) decreases with increasing SED-derived dust attenuation, the UV continuum slope, stellar mass and SFR. This suggests that Ly$\alpha$ preferentially escapes from blue galaxies with low dust attenuation, although a population of red LAEs is also present indicating that dust and Ly$\alpha$ are not mutually exclusive.
	
	\item By using completely different and independent methods to recover the total SFR, we obtain that the H$\alpha$ emission is an excellent star formation tracer at $z \sim 2$ with deviations typically lower than 0.3 dex for individual galaxies. These deviations are close to zero when averaging the sample in stellar mass or SFR bins.
		
	\item  By using the H$\alpha$-derived SFR we study the relation between SFR and stellar mass for our HAEs. We find a main-sequence (MS) of star formation, but with a slope lower than the classical \cite{Daddi2007} relation. By fitting a linear function in the form: $\log{\rm SFR} = a + b \times \log{\left(M / M_\odot \right)}$, we obtain: $a = -3.65 \pm 0.15$ and $b= 0.52 \pm 0.02$. We show that, in part, the lower slope with respect to previous works might be due to the different selection criteria. However, exploring the uncertainties in the slope and zero point of the HAE MS, we find that they are very sensitive to both the dust correction factors adopted to recover the total SFR and the way the stellar masses are determined. This largely explains the difference with previous works and represent the main uncertainty in the definition of MS at high redshift. This might apply to any sample of SF galaxies. Using the same recipes for stellar mass calculation and dust correction as in \cite{Daddi2007} we find consistent results for the MS at $z \sim 2$ for our HAEs and also for $sBzK$ galaxies whose total SFR are obtained with a stacking analysis in \emph{Herschel} bands. 
	
\end{enumerate}

\section*{Acknowledgments}

The authors acknowledge the anonymous referee for their detailed and constructive report which has improved the presentation of the results. IO and RJI acknowledge support from the European Research Council (ERC) in the form of Advanced Grant, {\sc cosmicism}. DS acknowledges financial support from the Netherlands Organisation for Scientific research (NWO) through a Veni fellowship, from FCT through a FCT Investigator Starting Grant and Start-up Grant (IF/01154/2012/CP0189/CT0010) and from FCT grant PEst-OE/FIS/UI2751/2014. IRS acknowledges support from STFC (ST/L00075X/1), the ERC Advanced Investigator programme DUSTYGAL 321334 and a Royal Society/Wolfson Merit Award. PNB acknowledges support from STFC. {\it Herschel} is an ESA space observatory with science instruments provided by European-led Principal Investigator consortia and with important participation from NASA. The {\it Herschel} spacecraft was designed, built, tested, and launched under a contract to ESA managed by the Herschel/Planck Project team by an industrial consortium under the overall responsibility of the prime contractor Thales Alenia Space (Cannes), and including Astrium (Friedrichshafen) responsible for the payload module and for system testing at spacecraft level, Thales Alenia Space (Turin) responsible for the service module, and Astrium (Toulouse) responsible for the telescope, with in excess of a hundred subcontractors. PACS has been developed by a consortium of institutes led by MPE (Germany) and including UVIE (Austria); KUL, CSL, IMEC (Belgium); CEA, OAMP (France); MPIA (Germany); IFSI, OAP/AOT, OAA/CAISMI, LENS, SISSA (Italy); IAC (Spain). This development has been supported by the funding agencies BMVIT (Austria), ESA-PRODEX (Belgium), CEA/CNES (France), DLR (Germany), ASI (Italy) and CICYT/MICINN (Spain). The HerMES data was accessed through the HeDaM database (http://hedam.oamp.fr) operated by CeSAM and hosted by the Laboratoire d'Astrophysique de Marseille.

\bibliographystyle{mn2e}

\bibliography{ioteo_biblio}

\newpage

\label{lastpage}

\end{document}